\documentclass[pra,aps,onecolumn,superscriptaddress,showpacs,showkeys,nofootinbib]{revtex4}
\usepackage{amsmath,amsthm,amssymb,graphicx,subfigure,xcolor}
\usepackage{mathrsfs}
\usepackage{graphics,float}
\usepackage{epstopdf}
\usepackage{color}
\usepackage[unicode=true]{hyperref}
\usepackage{booktabs}
\usepackage{tabularx}
\usepackage{tabu}
\usepackage{multirow}
\hypersetup{
     colorlinks=true,       		
     linkcolor=blue,          	
     citecolor=blue,            
     urlcolor=blue,           	
 }

\newtheorem*{theorem*}{Theorem}

\newtheorem*{corollary*}{Corollary}

\newtheorem*{lemma*}{Lemma}

\newtheorem*{proposition*}{Proposition}
\theoremstyle{definition}

\newtheorem*{definition*}{Definition}
\theoremstyle{remark}

\newtheorem*{remark*}{Remark}

\begin{document}

\title{Tighter sum uncertainty relations via metric-adjusted skew information}

\author{Hui Li}
\author{Ting Gao}
\email{gaoting@hebtu.edu.cn}
\affiliation{School of Mathematical Sciences, Hebei Normal University, Shijiazhuang 050024, China}
\author{Fengli Yan}
\email{flyan@hebtu.edu.cn}
\affiliation{College of Physics, Hebei Key Laboratory of Photophysics Research and Application, Hebei Normal University, Shijiazhuang 050024, China}
\begin{abstract}
In this paper, we first provide three general norm inequalities, which are used to give new uncertainty relations of any finite observables and quantum channels via metric-adjusted skew information. The results are applicable to its special cases as Wigner-Yanase-Dyson skew information. In quantifying the uncertainty of channels, we discuss two types of lower bounds and compare the tightness between them, meanwhile, a tight lower bound is given.  The uncertainty relations obtained by us are stronger than the existing ones. To illustrate our results, we give several specific examples.
\end{abstract}

\keywords{sum uncertainty relation, metric-adjusted skew information, observables, quantum channels}

\pacs{03.65.-w, 03.65.Ta}


\maketitle

\section{Introduction}
Uncertainty principle as a quintessential manifestation of quantum mechanics reveals the insights that distinguish quantum theory from classical theory. Heisenberg originally came up with the uncertainty principle \cite{100} in 1927, which enunciates that the position and momentum of a particle cannot be determined simultaneously. Since then, the quantitative characterization of the uncertainty relation has received extensive attention, and many results have been obtained.

There are a host of methods to characterize the uncertainty principle, one of which is variance. This method, which was adopted by Robertson \cite{101} and Schr\"{o}dinger \cite{102}, has been found that there exist lower bounds on the variances product for any two non-commuting observables. Subsequently, with regard to the sum of variance, the stronger uncertainty relations are provided \cite{103}. And Wang $et~al.$ \cite{104} verified the results in \cite{103} through experiment. Later, some tighter uncertainty relations with respect to  variance were given \cite{105,106,107,4}.

The other well-known method of characterizing uncertainty relation is entropy. Deutsch \cite{110} first proposed a quantitative expression of the uncertainty principle by entropy for any two non-commuting observables, and then Maassen and Uffink \cite{111} optimized the result in 1988. Furthermore, many scholars have put forward diverse uncertainty relations respecting distinct entropies \cite{112,121,113,114}. The uncertainty relations of entropy have numerous applications ranging from entanglement witnesses \cite{16,116}, quantum teleportation \cite{117}, quantum steering \cite{118}, quantum key distribution \cite{15,115} to quantum metrology \cite{119}.

Recently, Luo \cite{120} confirmed that skew information is an alternative approach to characterize the uncertainty relation. In \cite{122}, Wigner and Yanase introduced the definition of skew information
\begin{equation}\label{1}
\begin{aligned}
I_{\rho}(M)=-\frac{1}{2}{\rm{Tr}}[{\rho}^{{1}/{2}},M]^2=\frac{1}{2}\|[{\rho}^{{1}/{2}},M]\|^2.
\end{aligned}
\end{equation}
Here $\rho$ and $M$ represent the quantum state and the observable, respectively. It can be considered as a measure and quantifies the information content included in the state $\rho$ regarding the conserved observables. Meanwhile, compared to the usual variance, it is better at times. The skew information, for pure states, is the same as the variance \cite{125}, but they differ in mixed states. In the space of quantum states, skew information is convex, on the contrary for variance \cite{125}, which is one of the remarkable differences between them. Later, Dyson put forth a quantitative way which is a generalization of skew information, its specific expression is
\begin{equation}\label{2}
\begin{aligned}
I_{\rho}^{\alpha}(M)=-\frac{1}{2}{\rm Tr}[\rho^{\alpha},M][\rho^{1-\alpha},M],~~~~0<\alpha<1,\\
\end{aligned}
\end{equation}
termed as Wigner-Yanase-Dyson skew information, and Lieb \cite{123} resolved the convexity of this form on quantum states.

The sum of quantum uncertainty is crucial, because it is an effective tool for detecting quantum entanglement \cite{126,127,128,129,130}. To this end, the sum of quantum Fisher information (QFI) which was defined by means of symmetric logarithmic derivative probably is superior to Wigner-Yanase skew information \cite{128}, as in the quantum Cram\'{e}r-Rao inequality. In \cite{50}, Petz proposed the concept of monotone metric. After that, Hansen \cite{51} further developed the notion of monotone metric which is metric-adjusted skew information, and QFI can be considered as a particular case of it. Consequently, we would like to further study tighter uncertainty relations regarding metric-adjusted skew information.

Quantum channel is essential in quantum theory. The uncertainty relation of channels has also been investigated extensively, and a large number of results have been yielded \cite{2,5}. Recently, some scholars generalized the uncertainty inequalities to metric-adjusted skew information for arbitrary finite quantum channels \cite{1,3,27}.

The overall structure of this paper is as follows. In Sec. \ref{II}, we recall the notion of metric-adjusted skew information. In Sec. \ref{III}, we present some norm inequalities, and then new uncertainty relations of observables are given regarding metric-adjusted skew information. The distinct types of uncertainty relations of quantum channels as for metric-adjusted skew information are discussed in Sec. \ref{IV}, and we proved that which of the two corresponding lower bounds obtained by the same norm inequality is better. At the same time, the above conclusions still hold for its special metrics. We also give several examples and compare the lower bounds obtained by us with the lower bounds in \cite{3,1,27}. This more intuitively shows that our results are more accurate than the ones in \cite{3,1,27}. The main conclusions are summarized in Sec. \ref{V}.

\section{Metric-adjusted skew information}\label{II}
Suppose that $M_{n}(\mathbb{C})$ is the set of all complex $n\times n$ matrices, $\mathscr{M}_{n}$ is the set of all positive definite $n\times n$ matrices with trace 1, where $n\in\mathbb{N}$. For any $A,B\in M_{n}(\mathbb{C})$, $\rho\in\mathscr{M}_{n}$, $K_{\rho}(\cdot,\cdot):M_{n}(\mathbb{C})\times M_{n}(\mathbb{C})\mapsto\mathbb{C}$ is termed as symmetric monotone metric \cite{50} when it is content with the requirements

(i) $(A,B)\mapsto K_{\rho}(A,B)$ is sesquilinear, that is, the function $K_{\rho}(A,\cdot)$ is linear and $K_{\rho}(\cdot,B)$ conjugate linear.

(ii) $K_{\rho}(A,A)$ is nonnegative, $K_{\rho}(A,A)=0$ if and only if $A=0$.

(iii) $\rho\mapsto K_{\rho}(A,A)$ is continuous on $\mathscr{M}_{n}$.

(iv) $K_{T(\rho)}(T(A),T(A))\leq K_{\rho}(A,A)$ for any stochastic mapping $T$. A linear mapping $T:M_{n}(\mathbb{C})\rightarrow M_{m}(\mathbb{C})$ is called stochastic mapping if $T(\mathscr{M}_{n})\subset{\mathscr{M}_{n}}$ and $T$ is a completely positive.

(v) $K_{\rho}(A,B)=K_{\rho}(A^{\dagger},B^{\dagger})$.

The symmetric monotone metric $K_{\rho}(A,B)$ can be expressed as
\begin{equation}\label{3}
\begin{aligned}
K_{\rho}(A,B)={\rm{Tr}}[A^{\dagger}c(L_{\rho},R_{\rho})B],
\end{aligned}
\end{equation}
where $L_{\rho}$ and $R_{\rho}$ are respectively left and right multiplication operators, $c$ is termed as Morozova-Chentsov function, and its form is
\begin{equation}\label{4}
\begin{aligned}
c(x,y)=\frac{1}{yf(xy^{-1})},~~x,y>0,
\end{aligned}
\end{equation}
where the function $f$ is satisfied with the conditions: (a) $f:R_{+}\mapsto R_{+}$ is an operator monotone, where $R_{+}$ is the set of all positive real number, namely, if $A\geq B$, then $f(A)\geq f(B)$ for arbitrary $A,B>0$; (b) $tf(t^{-1})=f(t)$ for every $t>0$. Especially, it has been shown that if $K_{I/{n}}(I,I)=1$ holds, then the associated normalized function $f$ requires to admit $f(1)=1$. Here $I$ is $n$-dimensional identity operator.

In addition, in the space of quantum state, if the Morozova-Chentsov function associated with the symmetric monotone metric $K_{\rho}(\cdot,\cdot)$ satisfies
\begin{equation}\label{10}
\begin{aligned}
m(c)=\mathop{\mathrm{lim}}_{x\rightarrow0}c(x,1)^{-1}>0,
\end{aligned}
\end{equation}
then $K_{\rho}(\cdot,\cdot)$ is known as regular \cite{51}. $m(c)$ is called metric constant and $m(c)=f(0)$.

In \cite{51}, Hansen introduced the metric-adjusted skew information $I_{\rho}^{c}(M)$ which is
\begin{equation}\label{9}
\begin{aligned}
I_{\rho}^{c}(M)&=\frac{m(c)}{2}{K}_{\rho}^{c}({\rm i}[\rho,M],{\rm i}[\rho,M])\\
&=\frac{m(c)}{2}{\rm Tr}\left\{{\rm i}[\rho,M]c(L_{\rho},R_{\rho}){\rm i}[\rho,M]\right\},\\
\end{aligned}
\end{equation}
where $c$ satisfies the constraint (\ref{10}). Due to ${\rm i}[\rho,M]={\rm i}(L_{\rho}-R_{\rho})M$, then Eq. (\ref{9}) can be rewritten as
\begin{equation}
\begin{aligned}
I_{\rho}^{c}(M)=\frac{m(c)}{2}{\rm Tr}\left\{M\hat{c}(L_{\rho},R_{\rho})M\right\},\\
\end{aligned}
\end{equation}
where $\hat{c}(x,y)=(x-y)^2c(x,y)$, $x,y>0$.

When one chooses
\begin{equation}\label{5}
\begin{aligned}
c^{\rm{WY}}(x,y)=\frac{4}{(\sqrt{x}+\sqrt{y})^2}, ~~x,y>0,
\end{aligned}
\end{equation}
and
\begin{equation}\label{6}
\begin{aligned}
 c^{\alpha}(x,y)=\frac{1}{\alpha(1-\alpha)}\frac{(x^{\alpha}-y^{\alpha})(x^{1-\alpha}-y^{1-\alpha})}{(x-y)^2}, ~~0<\alpha<1,
\end{aligned}
\end{equation}
the associated monotone metrics
\begin{equation}\label{7}
\begin{aligned}
K_{\rho}^{\rm WY}(A,B)={\rm Tr}[A^{\dagger}c_{\rho}^{\rm WY}(L_{\rho},R_{\rho})B],
\end{aligned}
\end{equation}
and
\begin{equation}\label{8}
\begin{aligned}
K_{\rho}^{\alpha}(A,B)={\rm Tr}[A^{\dagger}c_{\rho}^{\alpha}(L_{\rho},R_{\rho})B]
\end{aligned}
\end{equation}
are known as Wigner-Yanase metric and Wigner-Yanase-Dyson metric, respectively. Therefore, when $c=c^\alpha$, Eq. (\ref{9}) turns into Eq. (\ref{2}) which is Wigner-Yanase-Dyson skew information $I_{\rho}^{\alpha}(M)$. When $\alpha=\frac{1}{2}$, Eq. (\ref{2}) reduces to Eq. (\ref{1}) which is Wigner-Yanase skew information $I_{\rho}(M)$.

\section{Uncertainty relations of finite observables}\label{III}
In this section, we first present some norm inequalities. By  using these  inequalities the new sum uncertainty relations of finite observables are given via metric-adjusted skew information, and the results also hold for its special metrics, such as those mentioned in Sec. \ref{II} above. Then we provide two examples which show that our results are better than existing ones.

For finite $n$ observables $M_{1},M_{2},\cdot\cdot\cdot, M_{n}~(n>2)$, Cai \cite{3} showed the uncertainty relation
\begin{equation}\label{11}
\begin{aligned}
\sum\limits_{i=1}^{n}{I}_{\rho}^{c}(M_{i})\geq\frac{1}{n-2}\Bigg[\sum\limits_{1\leq i<j\leq n}{I}_{\rho}^{c}(M_{i}+M_{j})-\frac{1}{(n-1)^2}\Bigg(\sum\limits_{1\leq i<j\leq n} \sqrt{{I}_{\rho}^{c}(M_{i}+M_{j})}\Bigg)^{2}\Bigg].\\
\end{aligned}
\end{equation}
Ren $et~al$. \cite{1} gave an uncertainty inequality
\begin{equation}\label{12}
\begin{aligned}
\sum\limits_{i=1}^{n}{I}_{\rho}^{c}(M_{i})\geq\frac{1}{n}{I}_{\rho}^{c}\Bigg(\sum\limits_{i=1}^{n}M_{i}\Bigg)+\frac{2}{n^2(n-1)}\Bigg(\sum\limits_{1\leq i<j\leq n} \sqrt{{I}_{\rho}^{c}(M_{i}-M_{j})}\Bigg)^{2}.\\
\end{aligned}
\end{equation}
Recently, Zhang $et~al$. \cite{27} provided an uncertainty inequality
\begin{equation}\label{40}
\begin{aligned}
&\sum\limits_{i=1}^{n}{I}_{\rho}^{c}(M_{i})\geq\max_{z\in\{0,1\}}\frac{1}{2n-2}{\left\{\frac{2}{n(n-1)}\Bigg(\sum\limits_{1\leq i<j\leq n} \sqrt{{I}_{\rho}^{c}(M_{i}+(-1)^z M_{j})}\Bigg)^{2}+\sum\limits_{1\leq i<j\leq n}{I}_{\rho}^{c}(M_{i}+(-1)^{z+1}M_{j})\right\}}.\\
\end{aligned}
\end{equation}
The inequalities (\ref{12}) and (\ref{40}) hold when $n\geq2$. For simplicity, the lower bounds in (\ref{11}), (\ref{12}), and (\ref{40}) are marked by $Lb_{1}$, $Lb_{2}$, and $Lb_{3}$, respectively.

Next we show various inequalities of the norm which are essential for the discussion of main results, so we take the inequalities as a Lemma.

\textbf{Lemma 1}. Suppose that ${\mathbf x}_{i}$ is a complex matrix, we can get
\begin{equation}\label{013}
\sum\limits_{i=1}^{n}\|{{\mathbf x}_{i}}\|^2\geq\frac {1}{mn+(n-2)l}\left[ {\frac{2l}{n(n-1)}\Bigg(\sum\limits_{1\leq i<j\leq n}\|{{\mathbf x}_{i}}+{{\mathbf x}_{j}}\|\Bigg)^2}+m\sum\limits_{1\leq i<j\leq n}\|{{\mathbf x}_{i}}-{{\mathbf x}_{j}}\|^2+(m-l)\Bigg\|\sum\limits_{i=1}^{n}{\mathbf x}_{i}\Bigg\|^2\right]\\
\end{equation}
and
\begin{equation}\label{014}
\sum\limits_{i=1}^{n}\|{{\mathbf x}_{i}}\|^2\geq\frac {1}{mn+(n-2)l}\left[l \sum\limits_{1\leq i<j\leq n}\|{{\mathbf x}_{i}}+{{\mathbf x}_{j}}\|^2+{\frac{2m}{n(n-1)}\Bigg(\sum\limits_{1\leq i<j\leq n}\|{{\mathbf x}_{i}}-{{\mathbf x}_{j}}\|\Bigg)^2}+(m-l)\Bigg\|\sum\limits_{i=1}^{n}{\mathbf x}_{i}\Bigg\|^2\right]\\
\end{equation}
for  arbitrary $m, l> 0$, and
\begin{equation}\label{60}
\sum\limits_{i=1}^{n}\|{{\mathbf x}_{i}}\|^2\geq\frac {1}{mn+(n-2)l}\left[l \sum\limits_{1\leq i<j\leq n}\|{{\mathbf x}_{i}}+{{\mathbf x}_{j}}\|^2+m\sum\limits_{1\leq i<j\leq n}\|{{\mathbf x}_{i}}-{{\mathbf x}_{j}}\|^2+\frac{m-l}{(n-1)^2}\Bigg(\sum\limits_{1\leq i<j\leq n}\|{{\mathbf x}_{i}}+{{\mathbf x}_{j}}\|\Bigg)^2\right]\\
\end{equation}
for $l>m>0$. Special we have
\begin{equation}\label{13}
\begin{aligned}
\sum\limits_{i=1}^{n}\|{{\mathbf x}_{i}}\|^2\geq\frac{1}{3n-2}\Bigg[\frac{2}{n(n-1)}\Bigg(\sum\limits_{1\leq i<j\leq n}\|{{\mathbf x}_{i}}+{{\mathbf x}_{j}}\|\Bigg)^2+2\sum\limits_{1\leq i<j\leq n}\|{{\mathbf x}_{i}}-{{\mathbf x}_{j}}\|^2+\Bigg\|\sum\limits_{i=1}^{n}{\mathbf x}_{i}\Bigg\|^2\Bigg],\\
\end{aligned}
\end{equation}
and
\begin{equation}\label{14}
\begin{aligned}
\sum\limits_{i=1}^{n}\|{{\mathbf x}_{i}}\|^2\geq\frac{1}{3n-4}\Bigg[2\sum\limits_{1\leq i<j\leq n}\|{{\mathbf x}_{i}}+{{\mathbf x}_{j}}\|^2+\frac{2}{n(n-1)}\Bigg(\sum\limits_{1\leq i<j\leq n}\|{{\mathbf x}_{i}}-{{\mathbf x}_{j}}\|\Bigg)^2-\Bigg\|\sum\limits_{i=1}^{n}{\mathbf x}_{i}\Bigg\|^2\Bigg],\\
\end{aligned}
\end{equation}
and
\begin{equation}\label{61}
\sum\limits_{i=1}^{n}\|{{\mathbf x}_{i}}\|^2\geq\frac {1}{3n-4}\left[2 \sum\limits_{1\leq i<j\leq n}\|{{\mathbf x}_{i}}+{{\mathbf x}_{j}}\|^2+\sum\limits_{1\leq i<j\leq n}\|{{\mathbf x}_{i}}-{{\mathbf x}_{j}}\|^2-\frac{1}{(n-1)^2}\Bigg(\sum\limits_{1\leq i<j\leq n}\|{{\mathbf x}_{i}}+{{\mathbf x}_{j}}\|\Bigg)^2\right],\\
\end{equation}
where $\|\cdot\|$ denotes the operator norm of a matrix.

\textbf{Proof}. By using the equations
\begin{equation}
\begin{aligned}
\sum\limits_{1\leq i<j\leq n}\|{{\mathbf x}_{i}}+{{\mathbf x}_{j}}\|^2=\Bigg\|\sum\limits_{i=1}^{n}{\mathbf x}_{i}\Bigg\|^2+(n-2)\sum\limits_{i=1}^{n}\|{{\mathbf x}_{i}}\|^2,
\end{aligned}
\end{equation}
and
\begin{equation}
\begin{aligned}
\sum\limits_{1\leq i<j\leq n}\|{{\mathbf x}_{i}}-{{\mathbf x}_{j}}\|^2=n\sum\limits_{i=1}^{n}\|{{\mathbf x}_{i}}\|^2-\Bigg\|\sum\limits_{i=1}^{n}{\mathbf x}_{i}\Bigg\|^2,
\end{aligned}
\end{equation}
we can derive that
\begin{equation}
\sum\limits_{i=1}^{n}\|{{\mathbf x}_{i}}\|^2=\frac {1}{mn+(n-2)l}\left[l \sum\limits_{1\leq i<j\leq n}\|{{\mathbf x}_{i}}+{{\mathbf x}_{j}}\|^2+m\sum\limits_{1\leq i<j\leq n}\|{{\mathbf x}_{i}}-{{\mathbf x}_{j}}\|^2+(m-l)\Bigg\|\sum\limits_{i=1}^{n}{\mathbf x}_{i}\Bigg\|^2\right]
\end{equation}
for arbitrary $m, l\neq 0$ holds.

Then according to the inequality relations
\begin{equation}
\begin{aligned}
\sum\limits_{1\leq i<j\leq n}\|{{\mathbf x}_{i}}\pm{{\mathbf x}_{j}}\|^2\geq{\frac{2}{n(n-1)}\Bigg(\sum\limits_{1\leq i<j\leq n}\|{{\mathbf x}_{i}}\pm{{\mathbf x}_{j}}\|\Bigg)^2},
\end{aligned}
\end{equation}
we can get
\begin{equation}
\sum\limits_{i=1}^{n}\|{{\mathbf x}_{i}}\|^2\geq\frac {1}{mn+(n-2)l}\left[ {\frac{2l}{n(n-1)}\Bigg(\sum\limits_{1\leq i<j\leq n}\|{{\mathbf x}_{i}}+{{\mathbf x}_{j}}\|\Bigg)^2}+m\sum\limits_{1\leq i<j\leq n}\|{{\mathbf x}_{i}}-{{\mathbf x}_{j}}\|^2+(m-l)\Bigg\|\sum\limits_{i=1}^{n}{\mathbf x}_{i}\Bigg\|^2\right]
\end{equation}
and
\begin{equation}
\sum\limits_{i=1}^{n}\|{{\mathbf x}_{i}}\|^2\geq\frac {1}{mn+(n-2)l}\left[l \sum\limits_{1\leq i<j\leq n}\|{{\mathbf x}_{i}}+{{\mathbf x}_{j}}\|^2+{\frac{2m}{n(n-1)}\Bigg(\sum\limits_{1\leq i<j\leq n}\|{{\mathbf x}_{i}}-{{\mathbf x}_{j}}\|\Bigg)^2}+(m-l)\Bigg\|\sum\limits_{i=1}^{n}{\mathbf x}_{i}\Bigg\|^2\right]
\end{equation}
for $m, l\geq 0$.
Due to $\Bigg\|\sum\limits_{i=1}^{n}{\mathbf x}_{i}\Bigg\|^2\leq\frac{1}{(n-1)^2}\Bigg(\sum\limits_{1\leq i<j\leq n}\|{{\mathbf x}_{i}}+{{\mathbf x}_{j}}\|\Bigg)^2$, when $l>m>0$, we have

\begin{equation}
\sum\limits_{i=1}^{n}\|{{\mathbf x}_{i}}\|^2\geq\frac {1}{mn+(n-2)l}\left[l \sum\limits_{1\leq i<j\leq n}\|{{\mathbf x}_{i}}+{{\mathbf x}_{j}}\|^2+m\sum\limits_{1\leq i<j\leq n}\|{{\mathbf x}_{i}}-{{\mathbf x}_{j}}\|^2+\frac{m-l}{(n-1)^2}\Bigg(\sum\limits_{1\leq i<j\leq n}\|{{\mathbf x}_{i}}+{{\mathbf x}_{j}}\|\Bigg)^2\right].\\
\end{equation}

For special case $m=2,~l=1$, we obtain inequality (\ref{13}). In the case $m=1,~l=2$, one gets inequalities (\ref{14}) and (\ref{61}). $\hfill\blacksquare$

When $n~(\geq2)$ is determined, the larger $m$ and the smaller $l$, the bigger right side of inequalities (\ref{013}) and (\ref{60}), the larger $l$ and the smaller $m$, the bigger right side of inequality (\ref{014}).

Note that for $m\geq l>0$ we have
\begin{equation}\label{034}
\begin{aligned}
\sum\limits_{i=1}^{n}\|{{\mathbf x}_{i}}\|^2&\geq\frac{1}{mn+(n-2)l}\Bigg[\frac{2l}{n(n-1)}\Bigg(\sum\limits_{1\leq i<j\leq n}\|{{\mathbf x}_{i}}+{{\mathbf x}_{j}}\|\Bigg)^2+m\sum\limits_{1\leq i<j\leq n}\|{{\mathbf x}_{i}}-{{\mathbf x}_{j}}\|^2+(m-l)\Bigg\|\sum\limits_{i=1}^{n}{\mathbf x}_{i}\Bigg\|^2\Bigg]\\
&\geq\frac{1}{2n-2}\Bigg[\frac{2}{n(n-1)}\Bigg(\sum\limits_{1\leq i<j\leq n}\|{{\mathbf x}_{i}}+{{\mathbf x}_{j}}\|\Bigg)^2+\sum\limits_{1\leq i<j\leq n}\|{{\mathbf x}_{i}}-{{\mathbf x}_{j}}\|^2\Bigg],\\
\end{aligned}
\end{equation}
and for $l\geq m>0$ one obtains
\begin{equation}\label{035}
\begin{aligned}
\sum\limits_{i=1}^{n}\|{\mathbf x}_{i}\|^2&\geq\frac{1}{mn+(n-2)l}\Bigg[\frac{2m}{n(n-1)}\Bigg(\sum\limits_{1\leq i<j\leq n}\|{{\mathbf x}_{i}}-{{\mathbf x}_{j}}\|\Bigg)^2+l\sum\limits_{1\leq i<j\leq n}\|{{\mathbf x}_{i}}+{{\mathbf x}_{j}}\|^2+(m-l)\Bigg\|\sum\limits_{i=1}^{n}{\mathbf x}_{i}\Bigg\|^2\Bigg]\\
&\geq\frac{1}{2n-2}\Bigg[\frac{2}{n(n-1)}\Bigg(\sum\limits_{1\leq i<j\leq n}\|{{\mathbf x}_{i}}-{{\mathbf x}_{j}}\|\Bigg)^2+\sum\limits_{1\leq i<j\leq n}\|{{\mathbf x}_{i}}+{{\mathbf x}_{j}}\|^2\Bigg]\\
&\geq\frac{1}{n}\Bigg\|\sum\limits_{i=1}^{n}{\mathbf x}_{i}\Bigg\|^2+\frac{2}{n^2(n-1)}\Bigg(\sum\limits_{1\leq i<j\leq n}\|{{\mathbf x}_{i}}-{{\mathbf x}_{j}}\|\Bigg)^2,\\
\end{aligned}
\end{equation}
and for $l> m>0$ we get
\begin{equation}\label{62}
\begin{aligned}
\sum\limits_{i=1}^{n}\|{{\mathbf x}_{i}}\|^2&\geq\frac {1}{mn+(n-2)l}\left[l \sum\limits_{1\leq i<j\leq n}\|{{\mathbf x}_{i}}+{{\mathbf x}_{j}}\|^2+m\sum\limits_{1\leq i<j\leq n}\|{{\mathbf x}_{i}}-{{\mathbf x}_{j}}\|^2+\frac{m-l}{(n-1)^2}\Bigg(\sum\limits_{1\leq i<j\leq n}\|{{\mathbf x}_{i}}+{{\mathbf x}_{j}}\|\Bigg)^2\right]\\
&\geq\frac {1}{n-2}\left[\sum\limits_{1\leq i<j\leq n}\|{{\mathbf x}_{i}}+{{\mathbf x}_{j}}\|^2-\frac{1}{(n-1)^2}\Bigg(\sum\limits_{1\leq i<j\leq n}\|{{\mathbf x}_{i}}+{{\mathbf x}_{j}}\|\Bigg)^2\right].\\
\end{aligned}
\end{equation}
Here the second inequality of (\ref{034}) is strictly greater than when $m>l>0$, the second inequality of (\ref{035}) and (\ref{62}) is strictly greater than when $l>m>0$ and $n>2$. So the inequality (\ref{13}) and the inequality in \cite{5} have the following relation
\begin{equation}\label{34}
\begin{aligned}
\sum\limits_{i=1}^{n}\|{{\mathbf x}_{i}}\|^2&\geq\frac{1}{3n-2}\Bigg[\frac{2}{n(n-1)}\Bigg(\sum\limits_{1\leq i<j\leq n}\|{{\mathbf x}_{i}}+{{\mathbf x}_{j}}\|\Bigg)^2+2\sum\limits_{1\leq i<j\leq n}\|{{\mathbf x}_{i}}-{{\mathbf x}_{j}}\|^2+\Bigg\|\sum\limits_{i=1}^{n}{\mathbf x}_{i}\Bigg\|^2\Bigg]\\
&>\frac{1}{2n-2}\Bigg[\frac{2}{n(n-1)}\Bigg(\sum\limits_{1\leq i<j\leq n}\|{{\mathbf x}_{i}}+{{\mathbf x}_{j}}\|\Bigg)^2+\sum\limits_{1\leq i<j\leq n}\|{{\mathbf x}_{i}}-{{\mathbf x}_{j}}\|^2\Bigg],\\
\end{aligned}
\end{equation}
and the inequality (\ref{14}) and the inequalities in \cite{2,5} have the relation
\begin{equation}\label{35}
\begin{aligned}
\sum\limits_{i=1}^{n}\|{\mathbf x}_{i}\|^2&\geq\frac{1}{3n-4}\Bigg[\frac{2}{n(n-1)}\Bigg(\sum\limits_{1\leq i<j\leq n}\|{{\mathbf x}_{i}}-{{\mathbf x}_{j}}\|\Bigg)^2+2\sum\limits_{1\leq i<j\leq n}\|{{\mathbf x}_{i}}+{{\mathbf x}_{j}}\|^2-\Bigg\|\sum\limits_{i=1}^{n}{\mathbf x}_{i}\Bigg\|^2\Bigg]\\
&\geq\frac{1}{2n-2}\Bigg[\frac{2}{n(n-1)}\Bigg(\sum\limits_{1\leq i<j\leq n}\|{{\mathbf x}_{i}}-{{\mathbf x}_{j}}\|\Bigg)^2+\sum\limits_{1\leq i<j\leq n}\|{{\mathbf x}_{i}}+{{\mathbf x}_{j}}\|^2\Bigg]\\
&\geq\frac{1}{n}\Bigg\|\sum\limits_{i=1}^{n}{\mathbf x}_{i}\Bigg\|^2+\frac{2}{n^2(n-1)}\Bigg(\sum\limits_{1\leq i<j\leq n}\|{{\mathbf x}_{i}}-{{\mathbf x}_{j}}\|\Bigg)^2,\\
\end{aligned}
\end{equation}
and when $n>2$, the relation between the inequality (\ref{61}) and the inequality in \cite{28} is

\begin{equation}\label{63}
\begin{aligned}
\sum\limits_{i=1}^{n}\|{{\mathbf x}_{i}}\|^2&\geq\frac {1}{3n-4}\left[2 \sum\limits_{1\leq i<j\leq n}\|{{\mathbf x}_{i}}+{{\mathbf x}_{j}}\|^2+\sum\limits_{1\leq i<j\leq n}\|{{\mathbf x}_{i}}-{{\mathbf x}_{j}}\|^2-\frac{1}{(n-1)^2}\Bigg(\sum\limits_{1\leq i<j\leq n}\|{{\mathbf x}_{i}}+{{\mathbf x}_{j}}\|\Bigg)^2\right]\\
&>\frac {1}{n-2}\left[\sum\limits_{1\leq i<j\leq n}\|{{\mathbf x}_{i}}+{{\mathbf x}_{j}}\|^2-\frac{1}{(n-1)^2}\Bigg(\sum\limits_{1\leq i<j\leq n}\|{{\mathbf x}_{i}}+{{\mathbf x}_{j}}\|\Bigg)^2\right].\\
\end{aligned}
\end{equation}

These inequalities are helpful for us to explore the tighter uncertainty relations. Based on the inequalities (\ref{013})---(\ref{61}), we give tighter sum uncertainty relations in the following Theorem.

\textbf{Theorem 1}. For finite $n$ observables $M_{1},M_{2},\cdot\cdot\cdot, M_{n}~(n\geq2)$, the sum uncertainty relation with respect to  metric-adjusted skew information is
\begin{equation}\label{17}
\begin{aligned}
\sum\limits_{i=1}^{n}{I}_{\rho}^{c}(M_{i})\geq\max\{\rm{ineq\ref{16},ineq\ref{15},ineq\ref{64}}\},\\
\end{aligned}
\end{equation}
Specially, we have
\begin{equation}
\begin{aligned}
\sum\limits_{i=1}^{n}{I}_{\rho}^{c}(M_{i})\geq\max\{\rm{ineq\ref{016},ineq\ref{015},ineq\ref{65}}\},\\
\end{aligned}
\end{equation}
where the ineq\ref{16}---ineq\ref{65} represent the lower bounds of inequality (\ref{16})---inequality (\ref{65}), respectively.

\textbf{Proof}. Because the symmetric monotone metrics $K_{\rho}^{c}(\cdot,\cdot)$ are satisfied with the norm property, according to inequality (\ref{013}), one has
\begin{equation}\label{018}
\begin{aligned}
\sum\limits_{i=1}^{n}K_{\rho}^{c}({\rm i}[\rho,M_{i}],&{\rm i}[\rho,M_{i}])\geq\frac {1}{mn+(n-2)l}{\left\{\frac{2l}{n(n-1)}
\Bigg(\sum\limits_{1\leq i<j\leq n} \sqrt{K_{\rho}^{c}({\rm i}[\rho,M_{i}+M_{j}],{\rm i}[\rho,M_{i}+M_{j}])}\Bigg)^{2}+\right.}\\
&{\left.m\sum\limits_{1\leq i<j\leq n}K_{\rho}^{c}({\rm i}[\rho,M_{i}-M_{j}],{\rm i}[\rho,M_{i}-M_{j}])+(m-l)K_{\rho}^{c}\bigg({\rm i}\bigg[\rho,\sum\limits_{i=1}^{n} M_{i}\bigg],{\rm i}\bigg[\rho,\sum\limits_{i=1}^{n} M_{i}\bigg]\bigg)\right\}}\\
\end{aligned}
\end{equation}
for $m\geq l>0$. Multiply both sides of inequality (\ref{018}) by a constant $\frac{f(0)}{2}$, we can derive
\begin{equation}\label{16}
\begin{aligned}
\sum\limits_{i=1}^{n}{I}_{\rho}^{c}(M_{i})\geq&\frac {1}{mn+(n-2)l}{\left\{\frac{2l}{n(n-1)}\Bigg(\sum\limits_{1\leq i<j\leq n} \sqrt{{I}_{\rho}^{c}(M_{i}+M_{j})}\Bigg)^{2}+\right.}\\
&{\left.m\sum\limits_{1\leq i<j\leq n}{I}_{\rho}^{c}(M_{i}-M_{j})+(m-l){I}_{\rho}^{c}\bigg(\sum\limits_{i=1}^{n} M_{i}\bigg)\right\}}.\\
\end{aligned}
\end{equation}
Using the similar procedure, for $l\geq m>0$ we can get
\begin{equation}\label{15}
\begin{aligned}
\sum\limits_{i=1}^{n}{I}_{\rho}^{c}(M_{i})\geq&\frac {1}{mn+(n-2)l}{\left\{\frac{2m}{n(n-1)}\Bigg(\sum\limits_{1\leq i<j\leq n} \sqrt{{I}_{\rho}^{c}(M_{i}-M_{j})}\Bigg)^{2}+\right.}\\
&{\left.l\sum\limits_{1\leq i<j\leq n}{I}_{\rho}^{c}(M_{i}+M_{j})+(m-l){I}_{\rho}^{c}\bigg(\sum\limits_{i=1}^{n}{M}_{i}\bigg)\right\}},\\
\end{aligned}
\end{equation}
and for $l> m>0$ one reads
\begin{equation}\label{64}
\begin{aligned}
\sum\limits_{i=1}^{n}{I}_{\rho}^{c}(M_{i})\geq&\frac {1}{mn+(n-2)l}{\left\{l\sum\limits_{1\leq i<j\leq n}{I}_{\rho}^{c}(M_{i}+M_{j})+m\sum\limits_{1\leq i<j\leq n}{I}_{\rho}^{c}(M_{i}-M_{j})\right.}\\
&{\left.+\frac{m-l}{(n-1)^2}\Bigg(\sum\limits_{1\leq i<j\leq n} \sqrt{{I}_{\rho}^{c}(M_{i}+M_{j})}\Bigg)^{2}\right\}}.\\
\end{aligned}
\end{equation}
If we take $m = 2$, $l = 1$, the inequality (\ref{16}) turns into
\begin{equation}\label{016}
\begin{aligned}
\sum\limits_{i=1}^{n}{I}_{\rho}^{c}(M_{i})\geq&\frac{1}{3n-2}{\left\{\frac{2}{n(n-1)}\Bigg(\sum\limits_{1\leq i<j\leq n} \sqrt{{I}_{\rho}^{c}(M_{i}+M_{j})}\Bigg)^{2}+2\sum\limits_{1\leq i<j\leq n}{I}_{\rho}^{c}(M_{i}-M_{j})+{I}_{\rho}^{c}\bigg(\sum\limits_{i=1}^{n} M_{i}\bigg)\right\}}.\\
\end{aligned}
\end{equation}
If we take $m =1$, $l=2$, the inequality (\ref{15}) becomes
\begin{equation}\label{015}
\begin{aligned}
\sum\limits_{i=1}^{n}{I}_{\rho}^{c}(M_{i})\geq&\frac{1}{3n-4}{\left\{\frac{2}{n(n-1)}\Bigg(\sum\limits_{1\leq i<j\leq n} \sqrt{{I}_{\rho}^{c}(M_{i}-M_{j})}\Bigg)^{2}+2\sum\limits_{1\leq i<j\leq n}{I}_{\rho}^{c}(M_{i}+M_{j})-{I}_{\rho}^{c}\bigg(\sum\limits_{i=1}^{n} M_{i}\bigg)\right\}},\\
\end{aligned}
\end{equation}
and the inequality (\ref{64}) reduces to
\begin{equation}\label{65}
\begin{aligned}
\sum\limits_{i=1}^{n}{I}_{\rho}^{c}(M_{i})\geq&\frac{1}{3n-4}{\left\{2\sum\limits_{1\leq i<j\leq n}{I}_{\rho}^{c}(M_{i}+M_{j})+\sum\limits_{1\leq i<j\leq n}{I}_{\rho}^{c}(M_{i}-M_{j})\right.}\\
&{\left.-\frac{1}{(n-1)^2}\Bigg(\sum\limits_{1\leq i<j\leq n} \sqrt{{I}_{\rho}^{c}(M_{i}+M_{j})}\Bigg)^{2}\right\}}.\\
\end{aligned}
\end{equation}

For convenience, the lower bound of formula (\ref{17}) is marked by $Lb$, that is, $Lb=\max\{\rm{ineq\ref{16},ineq\ref{15},ineq\ref{64}}\}$.  $\hfill\blacksquare$

The following we illustrate that the lower bound obtained by us is stronger than the lower bounds in \cite{3,1,27}.
Because the uncertainty relations are derived by using the norm inequalities, according to the relationship between the norm inequalities presented in (\ref{034}), (\ref{035}), and (\ref{62}), it is not difficult to show the lower bound $Lb$ obtained by us is more accurate than the lower bounds of inequalities (\ref{11}), (\ref{12}), and (\ref{40}).

It is acknowledged that different results can be obtained by taking different Morozova-Chentsov functions for metric-adjusted skew information. Herein, we first consider the Morozova-Chentsov function with the form of Eq. (\ref{6}). Meanwhile, the following results are obtained.

\textbf{Corollary 1}. For finite $n$ observables $M_{1},M_{2},\cdot\cdot\cdot, M_{n}~(n\geq2)$, the sum uncertainty relations with respect to Wigner-Yanase-Dyson skew information are that for $m\geq l>0$ we can obtain
\begin{equation}\label{19}
\begin{aligned}
\sum\limits_{i=1}^{n}{I}_{\rho}^{\alpha}(M_{i})\geq&\frac {1}{mn+(n-2)l}{\left\{\frac{2l}{n(n-1)}\Bigg(\sum\limits_{1\leq i<j\leq n} \sqrt{{I}_{\rho}^{\alpha}(M_{i}+M_{j})}\Bigg)^{2}+\right.}\\
&{\left.m\sum\limits_{1\leq i<j\leq n}{I}_{\rho}^{\alpha}(M_{i}-M_{j})+(m-l){I}_{\rho}^{\alpha}\bigg(\sum\limits_{i=1}^{n} M_{i}\bigg)\right\}},\\
\end{aligned}
\end{equation}
and for $l\geq m>0$ one derives
\begin{equation}\label{20}
\begin{aligned}
\sum\limits_{i=1}^{n}{I}_{\rho}^{\alpha}(M_{i})\geq&\frac {1}{mn+(n-2)l}{\left\{\frac{2m}{n(n-1)}\Bigg(\sum\limits_{1\leq i<j\leq n} \sqrt{{I}_{\rho}^{\alpha}(M_{i}-M_{j})}\Bigg)^{2}+\right.}\\
&{\left.l\sum\limits_{1\leq i<j\leq n}{I}_{\rho}^{\alpha}(M_{i}+M_{j})+(m-l){I}_{\rho}^{\alpha}\bigg(\sum\limits_{i=1}^{n}{M}_{i}\bigg)\right\}},\\
\end{aligned}
\end{equation}
and for $l> m>0$ one reads
\begin{equation}\label{66}
\begin{aligned}
\sum\limits_{i=1}^{n}{I}_{\rho}^{\alpha}(M_{i})\geq&\frac {1}{mn+(n-2)l}{\left\{l\sum\limits_{1\leq i<j\leq n}{I}_{\rho}^{\alpha}(M_{i}+M_{j})+m\sum\limits_{1\leq i<j\leq n}{I}_{\rho}^{\alpha}(M_{i}-M_{j})\right.}\\
&{\left.+\frac{m-l}{(n-1)^2}\Bigg(\sum\limits_{1\leq i<j\leq n} \sqrt{{I}_{\rho}^{\alpha}(M_{i}+M_{j})}\Bigg)^{2}\right\}}.\\
\end{aligned}
\end{equation}
Thus we have $\sum\limits_{i=1}^{n}{I}_{\rho}^{\alpha}(M_{i})\geq\max\{{\rm ineq}\ref{19},{\rm ineq}\ref{20},{\rm ineq}\ref{66}\}$, where ineq\ref{19}, ineq\ref{20}, and ineq\ref{66} represent the lower bounds of inequality (\ref{19}), inequality (\ref{20}), and inequality (\ref{66}), respectively.

When $\alpha=\frac{1}{2}$, the inequalities (\ref{19}), (\ref{20}), and (\ref{66}) can be further reduced to the uncertainty inequalities with respect to Wigner-Yanase skew information, as shown below. For $m\geq l>0$ we get

\begin{equation}\label{32}
\begin{aligned}
\sum\limits_{i=1}^{n}{I}_{\rho}(M_{i})\geq&\frac {1}{mn+(n-2)l}{\left\{\frac{2l}{n(n-1)}\Bigg(\sum\limits_{1\leq i<j\leq n} \sqrt{{I}_{\rho}(M_{i}+M_{j})}\Bigg)^{2}+\right.}\\
&{\left.m\sum\limits_{1\leq i<j\leq n}{I}_{\rho}(M_{i}-M_{j})+(m-l){I}_{\rho}\bigg(\sum\limits_{i=1}^{n} M_{i}\bigg)\right\}},\\
\end{aligned}
\end{equation}
and for $l\geq m>0$ one obtains
\begin{equation}\label{33}
\begin{aligned}
\sum\limits_{i=1}^{n}{I}_{\rho}(M_{i})\geq&\frac {1}{mn+(n-2)l}{\left\{\frac{2m}{n(n-1)}\Bigg(\sum\limits_{1\leq i<j\leq n} \sqrt{{I}_{\rho}(M_{i}-M_{j})}\Bigg)^{2}+\right.}\\
&{\left.l\sum\limits_{1\leq i<j\leq n}{I}_{\rho}(M_{i}+M_{j})+(m-l){I}_{\rho}\bigg(\sum\limits_{i=1}^{n}{M}_{i}\bigg)\right\}},\\
\end{aligned}
\end{equation}
and for $l> m>0$ one reads
\begin{equation}\label{67}
\begin{aligned}
\sum\limits_{i=1}^{n}{I}_{\rho}(M_{i})\geq&\frac {1}{mn+(n-2)l}{\left\{l\sum\limits_{1\leq i<j\leq n}{I}_{\rho}(M_{i}+M_{j})+m\sum\limits_{1\leq i<j\leq n}{I}_{\rho}(M_{i}-M_{j})\right.}\\
&{\left.+\frac{m-l}{(n-1)^2}\Bigg(\sum\limits_{1\leq i<j\leq n} \sqrt{{I}_{\rho}(M_{i}+M_{j})}\Bigg)^{2}\right\}}.\\
\end{aligned}
\end{equation}
So we have $\sum\limits_{i=1}^{n}{I}_{\rho}(M_{i})\geq\max\{{\rm ineq}\ref{32},{\rm ineq}\ref{33},{\rm ineq}\ref{67}\}$, where ineq\ref{32}, ineq\ref{33}, and ineq\ref{67} represent the lower bounds of inequality (\ref{32}), inequality (\ref{33}), and inequality (\ref{67}), respectively.

It is highly natural to get that the lower bound $\max\{\rm{ineq\ref{32},ineq\ref{33}},{\rm ineq}\ref{67}\}$ is superior to the lower bounds in \cite{4,2,29}.

Next we present two examples in term of Wigner-Yanase-Dyson skew information to illustrate the superiority of our result. In the examples below we consider a special case, where we take $m=2$, $l=1$ for inequality (\ref{16}), and $m=1$, $l=2$ for inequalities (\ref{15}) and (\ref{64}).

\begin{figure}[htbp]
\centering
\subfigure[]{\includegraphics[width=8cm,height=5cm]{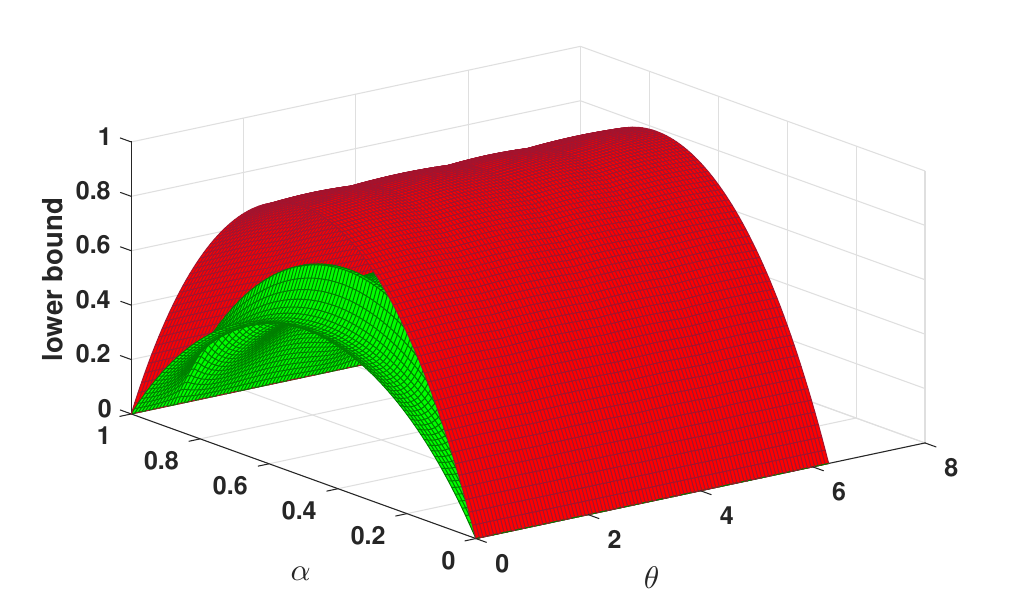}}
\quad
\subfigure[]{\includegraphics[width=8cm,height=5cm]{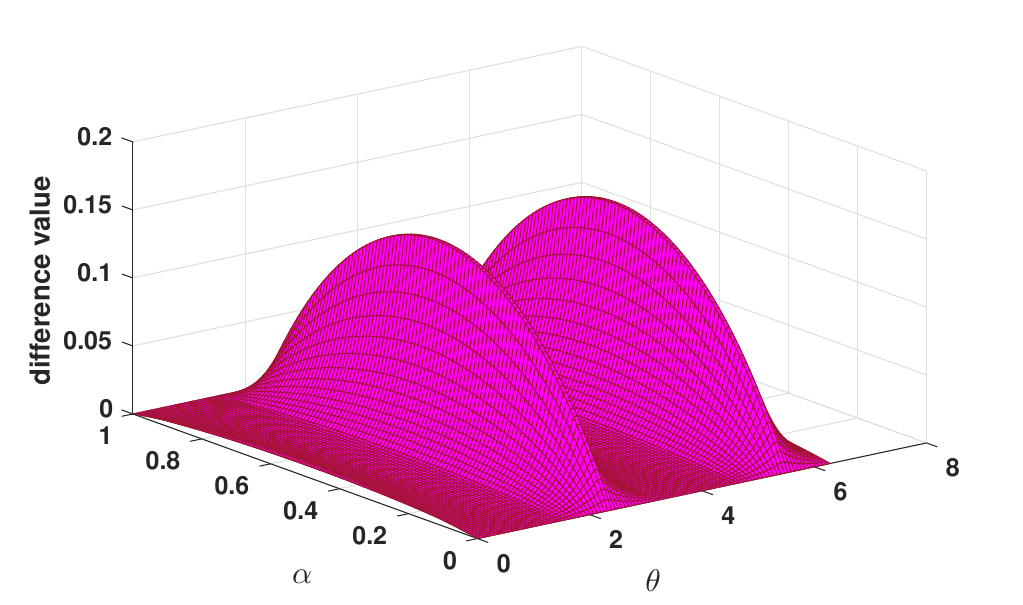}}
\quad
\subfigure[]{\includegraphics[width=8cm,height=5cm]{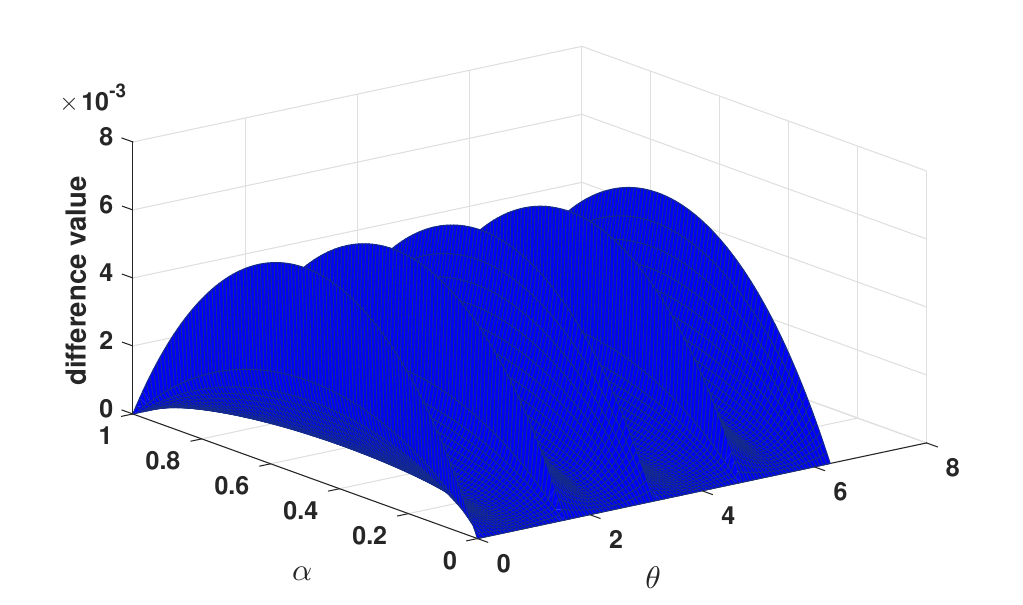}}
\quad
\subfigure[]{\includegraphics[width=8cm,height=5cm]{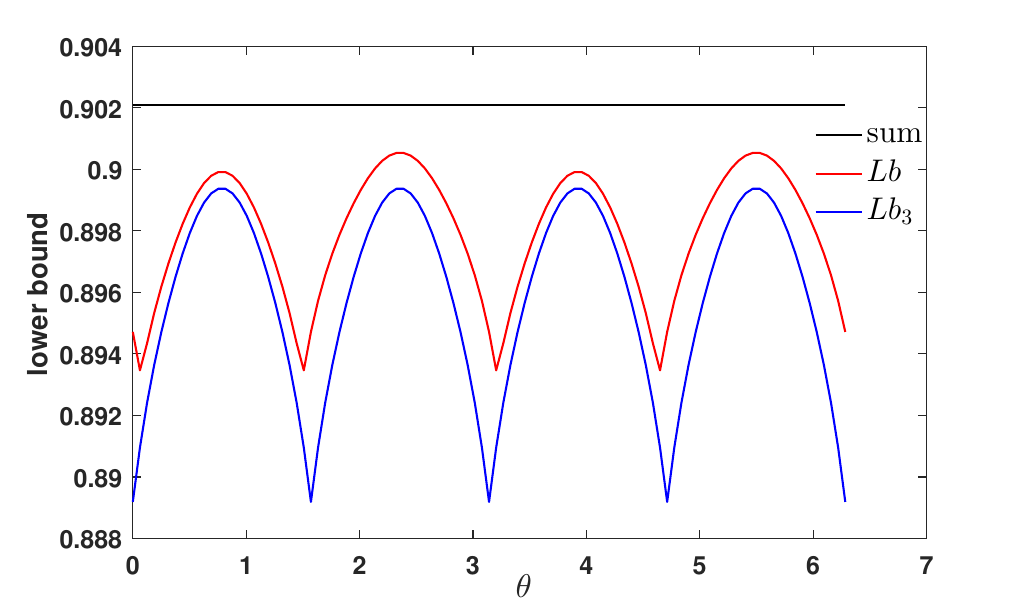}}
\caption{In (a), compared with the lower bounds for qubit states $\rho$. The red surface represents the lower bound $Lb$ for arbitrary $\alpha$; the green surface is the lower bound $Lb_{1}$ for arbitrary $\alpha$. The (b) stands for the difference value between the lower bound $Lb$ and $Lb_{2}$. The (c) denotes the difference value between the lower bound $Lb$ and $Lb_{3}$. Evidently, the lower bound $Lb$ is more accurate than $Lb_{1}$, $Lb_{2}$, $Lb_{3}$. Specially, in (d), we fix $\alpha=\frac{1}{3}$. The black line expresses ${I}_{\rho}^{1/3}(\sigma_{x})+{I}_{\rho}^{1/3}(\sigma_{y})+{I}_{\rho}^{1/3}(\sigma_{z})$; the red line and the blue line are the lower bounds $Lb$ and $Lb_{3}$, respectively.} \label{fig 1}
\end{figure}

\textbf{Example 1}. Assume a qubit state $\rho=\frac{I+\vec{r}\cdot\vec{\sigma}}{2}$ with $\vec{r}$=($\frac{\sqrt{3}}{2}$cos$\theta$, $\frac{\sqrt{3}}{2}$sin$\theta$, 0), and regard Pauli operators $\sigma_{x},~\sigma_{y},~\sigma_{z}$ as observables. The first three figures of FIG. \ref{fig 1} show the comparison of lower bounds for any $\alpha$. The (a) depicts the lower bounds $Lb$ and $Lb_{1}$. The difference value between the lower bound $Lb$ and $Lb_{2}$ is illustrated in (b), and $Lb-Lb_{2}$ is nonnegative, that is, $Lb\geq Lb_{2}$. Similarly, the lower bounds $Lb$ and $Lb_{3}$ are compared in (c), and $Lb\geq Lb_{3}$. Evidently, the lower bound $Lb$ is larger than $Lb_{1}$, $Lb_{2}$, $Lb_{3}$. Considering a special case, we take $\alpha=\frac{1}{3}$. In FIG. \ref{fig 1}(d), we only show the lower bounds $Lb$, $Lb_{3}$. And one can find that the lower bound obtained by us is closer to the sum ${I}_{\rho}^{1/3}(\sigma_{x})+{I}_{\rho}^{1/3}(\sigma_{y})+{I}_{\rho}^{1/3}(\sigma_{z})$.

\begin{figure}[htbp]
\centering
\subfigure[]{\includegraphics[width=8cm,height=5cm]{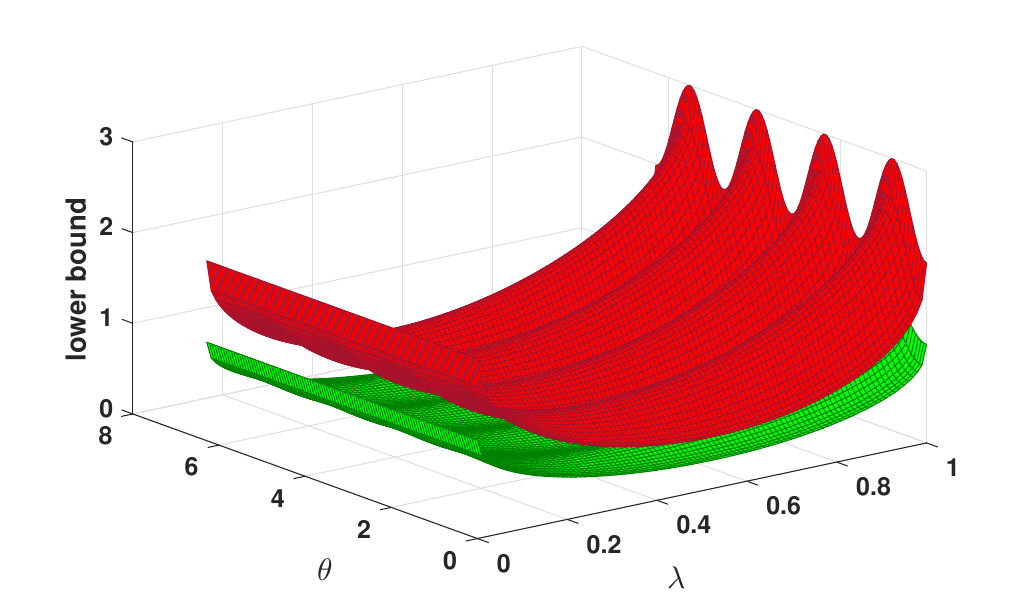}}
\quad
\subfigure[]{\includegraphics[width=8cm,height=5cm]{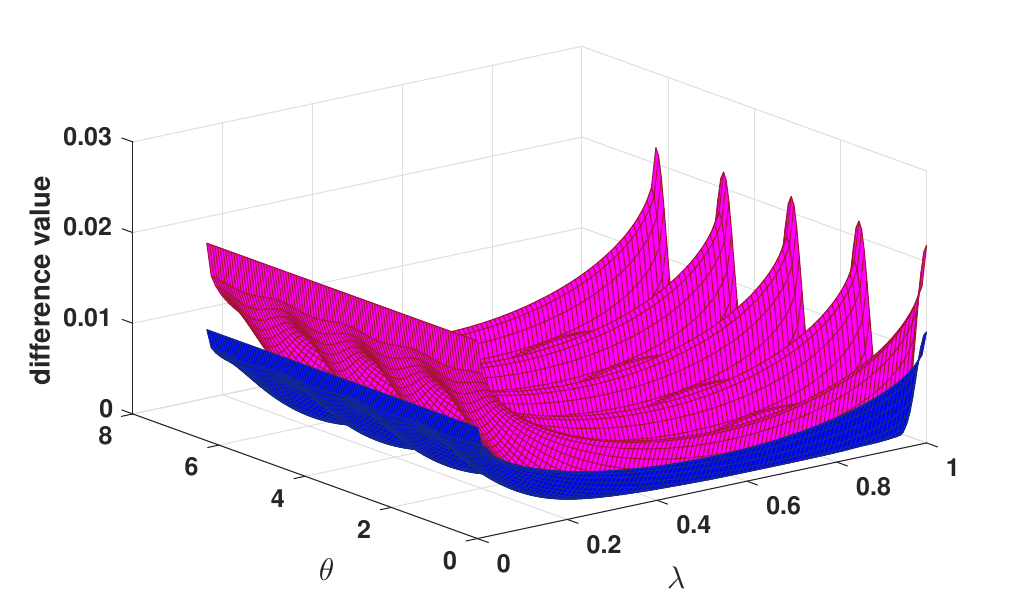}}
\caption{ In (a), compared with the lower bounds for Gisin state $\rho$, we fix $\alpha=\frac{1}{3}$. The red upper surface represents the lower bound $Lb$; the green surface stands for the lower bound $Lb_{1}$. Obviously, the lower bound $Lb$ we obtained is larger. The upper surface of (b) denotes the difference value between the lower bound $Lb$ and the lower bound $Lb_{2}$, the below surface of (b) is the difference value between the lower bound $Lb$ and the lower bound $Lb_{3}$.} \label{fig 2}
\end{figure}

\textbf{Example 2}. For a Gisin state $\rho=\lambda|\varphi(\theta)\rangle\langle\varphi(\theta)|+(1-\lambda)\sigma$ with $|\varphi(\theta)\rangle=\rm{sin}\theta|01\rangle-\rm{cos}\theta|10\rangle$, $\sigma=\frac{1}{2}|00\rangle\langle00|+\frac{1}{2}|11\rangle\langle11|$, $0\leq\lambda\leq1$, and $0\leq\theta\leq2\pi$. The operators $I\otimes\sigma_{x},~I\otimes\sigma_{y},~I\otimes\sigma_{z}$ are viewed as observables. Herein, we take $\alpha=\frac{1}{3}$. The FIG. \ref{fig 2}(a) depicts the lower bounds $Lb$ and $Lb_{1}$. The difference values of lower bounds are shown in FIG. \ref{fig 2}(b) which depicts $Lb-Lb_{3}$ and $Lb-Lb_{2}$, and they are nonnegative. Therefore, the lower bound $Lb$ obtained by us is the most accurate.

\section{Uncertainty relations of finite quantum channels}\label{IV}
In this section, the different types of uncertainty relations associated with any finite quantum channels are presented in terms of metric-adjusted skew information, and the conclusions also hold for its special metrics, such as those mentioned above in Sec. \ref{II}. In addition, we prove which of the two corresponding forms yields a better lower bound, and then an optimal lower bound is given. We also provide two examples for the sake of illustrating our results.

Given a quantum state $\rho$ and a quantum channel $\Phi$ represented by Kraus operators $\Phi(\rho)=\sum\limits_{j}K_{j}\rho K_{j}^{\dagger}$. In \cite{3}, Cai gave an uncertainty quantification associated with channel $\Phi$ with regard to metric-adjusted skew information
\begin{equation}\label{21}
\begin{aligned}
{I}_{\rho}^{c}(\Phi)=\sum\limits_{j}{I}_{\rho}^{c}(K_{j}),
\end{aligned}
\end{equation}
where
\begin{equation}\label{22}
\begin{aligned}
{I}_{\rho}^{c}(K_{j})&=\frac{m(c)}{2}{K}_{\rho}^{c}({\rm i}[\rho,K_{j}],{\rm i}[\rho,K_{j}])\\
&=\frac{m(c)}{2}{\rm Tr}\left\{{\rm i}[\rho,K_{j}^{\dagger}]c(L_{\rho},R_{\rho}){\rm i}[\rho,K_{j}]\right\}.
\end{aligned}
\end{equation}

Analogously, ${I}_{\rho}^{c}(\Phi)$ reduces to Wigner-Yanase-Dyson skew information ${I}_{\rho}^{\alpha}(\Phi)$ when $c=c^\alpha$, the specific form is
\begin{equation}\label{23}
\begin{aligned}
{I}_{\rho}^{\alpha}(\Phi)=\sum\limits_{j}{I}_{\rho}^{\alpha}(K_{j})=-\frac{1}{2}\sum\limits_{j}{\rm{Tr}}[\rho^\alpha,K_{j}^{\dagger}]
[\rho^{1-\alpha},K_{j}].
\end{aligned}
\end{equation}
When $\alpha=\frac{1}{2}$, ${I}_{\rho}^{\alpha}(\Phi)$ can be turned into the form
\begin{equation}\label{24}
\begin{aligned}
{I}_{\rho}(\Phi)=\sum\limits_{j}{I}_{\rho}(K_{j})=-\frac{1}{2}\sum\limits_{j}{\rm{Tr}}[{\rho}^{{1}/{2}},K_{j}^{\dagger}][{\rho}^{{1}/{2}},K_{j}],
\end{aligned}
\end{equation}
which is Wigner-Yanase skew information associated with channel.

For arbitrary $N$ quantum channels $\Phi_{1},\Phi_{2},\cdot\cdot\cdot, \Phi_{N}~(N\geq2)$, and each channel $\Phi_{t}$ is represented by Kraus operators, i.e., $\Phi_{t}(\rho)=\sum\limits_{j=1}^{n}K_{j}^{t}\rho(K_{j}^{t})^{\dagger}$, $t=1,2,\cdot\cdot\cdot, N$. In \cite{1}, Ren $et~al.$ gave two sum uncertainty quantification associated with channels,
\begin{equation}\label{25}
\begin{aligned}
\sum\limits_{t=1}^{N}{I}_{\rho}^{c}(\Phi_{t})\geq&\mathop{\rm{max}}\limits_{\pi_{t},\pi_{s}\in S_{n}}\frac{1}{N-2}
{\left\{\sum\limits_{1\leq t<s\leq N}\sum\limits_{j=1}^{n}{I}_{\rho}^{c}\Big(K_{\pi_{t}(j)}^{t}+K_{\pi_{s}(j)}^{s}\Big)-\right.}\\
&{\left.\frac{1}{(N-1)^2}\Bigg[\sum\limits_{j=1}^{n}\Bigg(\sum\limits_{1\leq t<s\leq N} \sqrt{{I}_{\rho}^{c}\Big(K_{\pi_{t}(j)}^{t}+K_{\pi_{s}(j)}^{s}\Big)}\Bigg)^{2}\Bigg]\right\}},\\
\end{aligned}
\end{equation}
and
\begin{equation}\label{26}
\begin{aligned}
&\sum\limits_{t=1}^{N}{I}_{\rho}^{c}(\Phi_{t})\geq\mathop{\rm{max}}\limits_{\pi_{t},\pi_{s}\in S_{n}}{\left\{\frac{1}{N}\sum\limits_{j=1}^{n}{I}_{\rho}^{c}\Bigg(\sum\limits_{t=1}^{N}K_{\pi_{t}(j)}^{t}\Bigg)+
\frac{2}{N^2(N-1)}\Bigg[\sum\limits_{j=1}^{n}\Bigg(\sum\limits_{1\leq t<s\leq N} \sqrt{{I}_{\rho}^{c}\Big(K_{\pi_{t}(j)}^{t}-K_{\pi_{s}(j)}^{s}\Big)}\Bigg)^{2}\Bigg]\right\}}.\\
\end{aligned}
\end{equation}
The formula (\ref{25}) can be used when $N>2$, while the formula (\ref{26}) can be used when $N\geq2$. For simplicity, the lower bounds in (\ref{25}) and (\ref{26}) are marked by $\overline{LB_{1}}$, and $\overline{LB_{2}}$, respectively.

Next, we will present the sum uncertainty relation of arbitrary finite $N$ quantum channels with regard to metric-adjusted skew information.

\textbf{Theorem 2}. For arbitrary $N$ quantum channels $\Phi_{1},\Phi_{2},\cdot\cdot\cdot, \Phi_{N}~(N\geq2)$, and each channel $\Phi_{t}$ is represented by Kraus operators, $\Phi_{t}(\rho)=\sum\limits_{j=1}^{n}K_{j}^{t}\rho(K_{j}^{t})^{\dagger}$, $t=1,2,\cdot\cdot\cdot, N$, one reads
\begin{equation}
\begin{aligned}
\sum\limits_{t=1}^{N}{I}_{\rho}^{c}(\Phi_{t})\geq\max\{\rm ineq\ref{27},ineq\ref{28},ineq\ref{68}\}.\\
\end{aligned}
\end{equation}
Specially, we have
\begin{equation}
\begin{aligned}
\sum\limits_{t=1}^{N}{I}_{\rho}^{c}(\Phi_{t})\geq\max\{\rm ineq\ref{027},ineq\ref{028},ineq\ref{69}\}.\\
\end{aligned}
\end{equation}
where ineq\ref{27}---ineq\ref{69} represent the lower bounds of inequality (\ref{27})---inequality (\ref{69}), respectively.

\textbf{Proof}. According to the norm inequality of Lemma 1, we can get
\begin{equation}\label{29}
\begin{aligned}
\sum\limits_{t=1}^{N}{I}_{\rho}^{c}(K_{\pi_{t}(j)}^{t})\geq &\frac{1}{mn+(n-2)l}{\left\{\frac{2l}{N(N-1)}\Bigg(\sum\limits_{1\leq t<s\leq N} \sqrt{{I}_{\rho}^{c}\Big(K_{\pi_{t}(j)}^{t}+K_{\pi_{s}(j)}^{s}\Big)}\Bigg)^{2}\right.}\\
&{\left.+m\sum\limits_{1\leq t<s\leq N}{I}_{\rho}^{c}\Big(K_{\pi_{t}(j)}^{t}-K_{\pi_{s}(j)}^{s}\Big)+(m-l){I}_{\rho}^{c}\Bigg(\sum\limits_{t=1}^{N}K_{\pi_{t}(j)}^{t}\Bigg)\right\}},\\
\end{aligned}
\end{equation}
both sides of this formula sum over $j$, for $m\geq l>0$ we have
\begin{equation}\label{27}
\begin{aligned}
\sum\limits_{t=1}^{N}{I}_{\rho}^{c}(\Phi_{t})\geq &\mathop{\rm{max}}\limits_{\pi_{t},\pi_{s}\in S_{n}}\frac{1}{mn+(n-2)l}{\left\{\frac{2l}{N(N-1)}\Bigg[\sum\limits_{j=1}^{n}\Bigg(\sum\limits_{1\leq t<s\leq N} \sqrt{{I}_{\rho}^{c}\Big(K_{\pi_{t}(j)}^{t}+K_{\pi_{s}(j)}^{s}\Big)}\Bigg)^{2}\Bigg]\right.}\\
&{\left.+m\sum\limits_{1\leq t<s\leq N}\sum\limits_{j=1}^{n}{I}_{\rho}^{c}\Big(K_{\pi_{t}(j)}^{t}-K_{\pi_{s}(j)}^{s}\Big)+(m-l)\sum\limits_{j=1}^{n}{I}_{\rho}^{c}
\Bigg(\sum\limits_{t=1}^{N}K_{\pi_{t}(j)}^{t}\Bigg)\right\}}.\\
\end{aligned}
\end{equation}
Using the similar method, for $l\geq m>0$ we can get
\begin{equation}\label{28}
\begin{aligned}
\sum\limits_{t=1}^{N}{I}_{\rho}^{c}(\Phi_{t})\geq &\mathop{\max}\limits_{\pi_{t},\pi_{s}\in S_{n}}\frac{1}{mn+(n-2)l}{\left\{\frac{2m}{N(N-1)}\Bigg[\sum\limits_{j=1}^{n}\Bigg(\sum\limits_{1\leq t<s\leq N} \sqrt{{I}_{\rho}^{c}\Big(K_{\pi_{t}(j)}^{t}-K_{\pi_{s}(j)}^{s}\Big)}\Bigg)^{2}\Bigg]\right.}\\
&{\left.+l\sum\limits_{1\leq t<s\leq N}\sum\limits_{j=1}^{n}{I}_{\rho}^{c}\Big(K_{\pi_{t}(j)}^{t}+K_{\pi_{s}(j)}^{s}\Big)+(m-l)\sum\limits_{j=1}^{n}{I}_{\rho}^{c}
\Bigg(\sum\limits_{t=1}^{N}K_{\pi_{t}(j)}^{t}\Bigg)\right\}},\\
\end{aligned}
\end{equation}
and for $l> m>0$ one obtains
\begin{equation}\label{68}
\begin{aligned}
\sum\limits_{t=1}^{N}{I}_{\rho}^{c}(\Phi_{t})\geq &\mathop{\rm{max}}\limits_{\pi_{t},\pi_{s}\in S_{n}}\frac{1}{mn+(n-2)l}{\left\{l\sum\limits_{1\leq t<s\leq N}\sum\limits_{j=1}^{n}{I}_{\rho}^{c}\Big(K_{\pi_{t}(j)}^{t}+K_{\pi_{s}(j)}^{s}\Big)+m\sum\limits_{1\leq t<s\leq N}\sum\limits_{j=1}^{n}{I}_{\rho}^{c}\Big(K_{\pi_{t}(j)}^{t}-K_{\pi_{s}(j)}^{s}\Big)\right.}\\
&{\left.+\frac{m-l}{(n-1)^2}\Bigg[\sum\limits_{j=1}^{n}\Bigg(\sum\limits_{1\leq t<s\leq N} \sqrt{{I}_{\rho}^{c}\Big(K_{\pi_{t}(j)}^{t}+K_{\pi_{s}(j)}^{s}\Big)}\Bigg)^{2}\Bigg]\right\}}.\\
\end{aligned}
\end{equation}
Specially, if we take $m=2$, $l=1$, the inequality (\ref{27}) turns into
\begin{equation}\label{027}
\begin{aligned}
\sum\limits_{t=1}^{N}{I}_{\rho}^{c}(\Phi_{t})\geq &\mathop{\rm{max}}\limits_{\pi_{t},\pi_{s}\in S_{n}}\frac{1}{3N-2}{\left\{\frac{2}{N(N-1)}\Bigg[\sum\limits_{j=1}^{n}\Bigg(\sum\limits_{1\leq t<s\leq N} \sqrt{{I}_{\rho}^{c}\Big(K_{\pi_{t}(j)}^{t}+K_{\pi_{s}(j)}^{s}\Big)}\Bigg)^{2}\Bigg]\right.}\\
&{\left.+2\sum\limits_{1\leq t<s\leq N}\sum\limits_{j=1}^{n}{I}_{\rho}^{c}\Big(K_{\pi_{t}(j)}^{t}-K_{\pi_{s}(j)}^{s}\Big)+\sum\limits_{j=1}^{n}{I}_{\rho}^{c}
\Bigg(\sum\limits_{t=1}^{N}K_{\pi_{t}(j)}^{t}\Bigg)\right\}},\\
\end{aligned}
\end{equation}
If one takes $m = 1$, $l = 2$, the inequalities (\ref{28}) and  (\ref{68}) respectively reduces into
\begin{equation}\label{028}
\begin{aligned}
\sum\limits_{t=1}^{N}{I}_{\rho}^{c}(\Phi_{t})\geq &\mathop{\rm{max}}\limits_{\pi_{t},\pi_{s}\in S_{n}}\frac{1}{3N-4}{\left\{\frac{2}{N(N-1)}\Bigg[\sum\limits_{j=1}^{n}\Bigg(\sum\limits_{1\leq t<s\leq N} \sqrt{{I}_{\rho}^{c}\Big(K_{\pi_{t}(j)}^{t}-K_{\pi_{s}(j)}^{s}\Big)}\Bigg)^{2}\Bigg]\right.}\\
&{\left.+2\sum\limits_{1\leq t<s\leq N}\sum\limits_{j=1}^{n}{I}_{\rho}^{c}\Big(K_{\pi_{t}(j)}^{t}+K_{\pi_{s}(j)}^{s}\Big)-\sum\limits_{j=1}^{n}{I}_{\rho}^{c}
\Bigg(\sum\limits_{t=1}^{N}K_{\pi_{t}(j)}^{t}\Bigg)\right\}}.\\
\end{aligned}
\end{equation}
and
\begin{equation}\label{69}
\begin{aligned}
\sum\limits_{t=1}^{N}{I}_{\rho}^{c}(\Phi_{t})\geq &\mathop{\rm{max}}\limits_{\pi_{t},\pi_{s}\in S_{n}}\frac{1}{3N-4}{\left\{2\sum\limits_{1\leq t<s\leq N}\sum\limits_{j=1}^{n}{I}_{\rho}^{c}\Big(K_{\pi_{t}(j)}^{t}+K_{\pi_{s}(j)}^{s}\Big)+\sum\limits_{1\leq t<s\leq N}\sum\limits_{j=1}^{n}{I}_{\rho}^{c}\Big(K_{\pi_{t}(j)}^{t}-K_{\pi_{s}(j)}^{s}\Big)\right.}\\
&{\left.-\frac{1}{(n-1)^2}\Bigg[\sum\limits_{j=1}^{n}\Bigg(\sum\limits_{1\leq t<s\leq N} \sqrt{{I}_{\rho}^{c}\Big(K_{\pi_{t}(j)}^{t}+K_{\pi_{s}(j)}^{s}\Big)}\Bigg)^{2}\Bigg]\right\}}.\\
\end{aligned}
\end{equation}
Here $\pi_{t},\pi_{s}\in S_{n}$ are $n$-element permutations. $\hfill\blacksquare$

By means of the norm inequality (\ref{035}), for $l\geq m>0$ we can derive
\begin{equation}
\begin{aligned}
\sum\limits_{t=1}^{N}{I}_{\rho}^{c}(K_{\pi_{t}(j)}^{t})\geq &\frac{1}{mn+(n-2)l}{\left\{\frac{2m}{N(N-1)}\Bigg(\sum\limits_{1\leq t<s\leq N} \sqrt{{I}_{\rho}^{c}\Big(K_{\pi_{t}(j)}^{t}-K_{\pi_{s}(j)}^{s}\Big)}\Bigg)^{2}\right.}\\
&{\left.+l\sum\limits_{1\leq t<s\leq N}{I}_{\rho}^{c}\Big(K_{\pi_{t}(j)}^{t}+K_{\pi_{s}(j)}^{s}\Big)+(m-l){I}_{\rho}^{c}\Bigg(\sum\limits_{t=1}^{N}K_{\pi_{t}(j)}^{t}\Bigg)\right\}}\\
\geq&{\frac{1}{N}{I}_{\rho}^{c}\Bigg(\sum\limits_{t=1}^{N}K_{\pi_{t}(j)}^{t}\Bigg)+
\frac{2}{N^2(N-1)}\Bigg(\sum\limits_{1\leq t<s\leq N} \sqrt{{I}_{\rho}^{c}\Big(K_{\pi_{t}(j)}^{t}-K_{\pi_{s}(j)}^{s}\Big)}\Bigg)^{2}},\\
\end{aligned}
\end{equation}
which leads to the result obtained by us being more accurate than the lower bound of inequality (\ref{26}). In the same way, we can also demonstrate that ineq\ref{68} is greater than $\overline{LB_1}$ based on inequality (\ref{62}).

The above results can be appropriate for its special measures, as Wigner-Yanase-Dyson skew information, thus the conclusions can be drawn as follows.

\textbf{Corollary 2}. For arbitrary $N$ quantum channels $\Phi_{1},\Phi_{2},\cdot\cdot\cdot, \Phi_{N}~(N\geq2)$, and each channel $\Phi_{t}$ is represented by Kraus operators, $\Phi_{t}(\rho)=\sum\limits_{j=1}^{n}K_{j}^{t}\rho(K_{j}^{t})^{\dagger}$, $t=1,2,\cdot\cdot\cdot, N$, for $m\geq l>0$ one has
\begin{equation}\label{30}
\begin{aligned}
\sum\limits_{t=1}^{N}{I}_{\rho}^{\alpha}(\Phi_{t})\geq &\mathop{\rm{max}}\limits_{\pi_{t},\pi_{s}\in S_{n}}\frac{1}{mn+(n-2)l}{\left\{\frac{2l}{N(N-1)}\Bigg[\sum\limits_{j=1}^{n}\Bigg(\sum\limits_{1\leq t<s\leq N} \sqrt{{I}_{\rho}^{\alpha}\Big(K_{\pi_{t}(j)}^{t}+K_{\pi_{s}(j)}^{s}\Big)}\Bigg)^{2}\Bigg]\right.}\\
&{\left.+m\sum\limits_{1\leq t<s\leq N}\sum\limits_{j=1}^{n}{I}_{\rho}^{\alpha}\Big(K_{\pi_{t}(j)}^{t}-K_{\pi_{s}(j)}^{s}\Big)+(m-l)\sum\limits_{j=1}^{n}{I}_{\rho}^{\alpha}
\Bigg(\sum\limits_{t=1}^{N}K_{\pi_{t}(j)}^{t}\Bigg)\right\}},\\
\end{aligned}
\end{equation}
and for $l\geq m>0$ we have
\begin{equation}\label{31}
\begin{aligned}
\sum\limits_{t=1}^{N}{I}_{\rho}^{\alpha}(\Phi_{t})\geq &\mathop{\rm{max}}\limits_{\pi_{t},\pi_{s}\in S_{n}}\frac{1}{mn+(n-2)l}{\left\{\frac{2m}{N(N-1)}\Bigg[\sum\limits_{j=1}^{n}\Bigg(\sum\limits_{1\leq t<s\leq N} \sqrt{{I}_{\rho}^{\alpha}\Big(K_{\pi_{t}(j)}^{t}-K_{\pi_{s}(j)}^{s}\Big)}\Bigg)^{2}\Bigg]\right.}\\
&{\left.+l\sum\limits_{1\leq t<s\leq N}\sum\limits_{j=1}^{n}{I}_{\rho}^{\alpha}\Big(K_{\pi_{t}(j)}^{t}+K_{\pi_{s}(j)}^{s}\Big)+(m-l)\sum\limits_{j=1}^{n}{I}_{\rho}^{\alpha}
\Bigg(\sum\limits_{t=1}^{N}K_{\pi_{t}(j)}^{t}\Bigg)\right\}},\\
\end{aligned}
\end{equation}
and for $l> m>0$ one obtains
\begin{equation}\label{70}
\begin{aligned}
\sum\limits_{t=1}^{N}{I}_{\rho}^{\alpha}(\Phi_{t})\geq &\mathop{\rm{max}}\limits_{\pi_{t},\pi_{s}\in S_{n}}\frac{1}{mn+(n-2)l}{\left\{l\sum\limits_{1\leq t<s\leq N}\sum\limits_{j=1}^{n}{I}_{\rho}^{\alpha}\Big(K_{\pi_{t}(j)}^{t}+K_{\pi_{s}(j)}^{s}\Big)+m\sum\limits_{1\leq t<s\leq N}\sum\limits_{j=1}^{n}{I}_{\rho}^{\alpha}\Big(K_{\pi_{t}(j)}^{t}\right.}\\
&{\left.-K_{\pi_{s}(j)}^{s}\Big)+\frac{m-l}{(n-1)^2}\Bigg[\sum\limits_{j=1}^{n}\Bigg(\sum\limits_{1\leq t<s\leq N} \sqrt{{I}_{\rho}^{\alpha}\Big(K_{\pi_{t}(j)}^{t}+K_{\pi_{s}(j)}^{s}\Big)}\Bigg)^{2}\Bigg]\right\}}.\\
\end{aligned}
\end{equation}
Therefore, $\sum\limits_{t=1}^{N}{I}_{\rho}^{\alpha}(\Phi_{t})\geq\max\{\rm ineq\ref{30},ineq\ref{31},ineq\ref{70}\}$, where ineq\ref{30}, ineq\ref{31}, and ineq\ref{70} represent the lower bounds of inequality (\ref{30}), inequality (\ref{31}), and inequality (\ref{70}), respectively.

When $\alpha=\frac{1}{2}$, the three uncertainty inequalities of Corollary 2 can be further simplified to Wigner-Yanase skew information, it is highly obvious that the lower bound of inequality (\ref{31}) is superior to the lower bound of \cite[Theorem 3]{2} according to the inequality (\ref{035}), and the lower bound of inequality (\ref{70}) is more precise than the lower bound of \cite[Theorem 2]{2} according to the inequality (\ref{62}).

The uncertainty quantification of channel $\Phi$ with regard to metric-adjusted skew information can also be expressed in the form
\begin{equation}\label{47}
\begin{aligned}
{I}_{\rho}^{c}(\Phi)&=\frac{m(c)}{2}{\rm Tr}\bigg\{\sum_{j=1}^n{\rm i}[\rho,K_{j}^{\dagger}]c(L_{\rho},R_{\rho}){\rm i}[\rho,K_{j}]\bigg\}\\
&=\frac{m(c)}{2}{\rm Tr}({\boldsymbol{\alpha}}^\dag I_{n}\otimes c(L_{\rho},R_{\rho}){\boldsymbol{\alpha}}).
\end{aligned}
\end{equation}
Here ${\boldsymbol{\alpha}}^\dag=({\rm i}[\rho,K_{1}^{\dagger}],\cdots,{\rm i}[\rho,K_{n}^{\dagger}])$. Therefore, on the basis of the inequalities  (\ref{013}), (\ref{014}), and (\ref{60}), for $m\geq l>0$ we have uncertainty relation
\begin{equation}\label{48}
\begin{aligned}
\sum\limits_{t=1}^{N}{I}_{\rho}^{c}(\Phi_{t})\geq &\mathop{\rm{max}}\limits_{\pi_{t},\pi_{s}\in S_{n}}\frac{1}{mn+(n-2)l}{\left\{\frac{2l}{N(N-1)}\Bigg(\sum\limits_{1\leq t<s\leq N} \sqrt{\sum\limits_{j=1}^{n}{I}_{\rho}^{c}\Big(K_{\pi_{t}(j)}^{t}+K_{\pi_{s}(j)}^{s}\Big)}\Bigg)^{2}\right.}\\
&{\left.+m\sum\limits_{1\leq t<s\leq N}\sum\limits_{j=1}^{n}{I}_{\rho}^{c}\Big(K_{\pi_{t}(j)}^{t}-K_{\pi_{s}(j)}^{s}\Big)+(m-l)\sum\limits_{j=1}^{n}{I}_{\rho}^{c}
\Bigg(\sum\limits_{t=1}^{N}K_{\pi_{t}(j)}^{t}\Bigg)\right\}},\\
\end{aligned}
\end{equation}
and for $l\geq m>0$ one reads
\begin{equation}\label{49}
\begin{aligned}
\sum\limits_{t=1}^{N}{I}_{\rho}^{c}(\Phi_{t})\geq &\mathop{\rm{max}}\limits_{\pi_{t},\pi_{s}\in S_{n}}\frac{1}{mn+(n-2)l}{\left\{\frac{2m}{N(N-1)}\Bigg(\sum\limits_{1\leq t<s\leq N} \sqrt{\sum\limits_{j=1}^{n}{I}_{\rho}^{c}\Big(K_{\pi_{t}(j)}^{t}-K_{\pi_{s}(j)}^{s}\Big)}\Bigg)^{2}\right.}\\
&{\left.+l\sum\limits_{1\leq t<s\leq N}\sum\limits_{j=1}^{n}{I}_{\rho}^{c}\Big(K_{\pi_{t}(j)}^{t}+K_{\pi_{s}(j)}^{s}\Big)+(m-l)\sum\limits_{j=1}^{n}{I}_{\rho}^{c}
\Bigg(\sum\limits_{t=1}^{N}K_{\pi_{t}(j)}^{t}\Bigg)\right\}},\\
\end{aligned}
\end{equation}
and for $l> m>0$ one derives
\begin{equation}\label{71}
\begin{aligned}
\sum\limits_{t=1}^{N}{I}_{\rho}^{c}(\Phi_{t})\geq &\mathop{\rm{max}}\limits_{\pi_{t},\pi_{s}\in S_{n}}\frac{1}{mn+(n-2)l}{\left\{l\sum\limits_{1\leq t<s\leq N}\sum\limits_{j=1}^{n}{I}_{\rho}^{c}\Big(K_{\pi_{t}(j)}^{t}+K_{\pi_{s}(j)}^{s}\Big)+m\sum\limits_{1\leq t<s\leq N}\sum\limits_{j=1}^{n}{I}_{\rho}^{c}\Big(K_{\pi_{t}(j)}^{t}-K_{\pi_{s}(j)}^{s}\Big)\right.}\\
&{\left.+\frac{m-l}{(n-1)^2}\Bigg(\sum\limits_{1\leq t<s\leq N} \sqrt{\sum\limits_{j=1}^{n}{I}_{\rho}^{c}\Big(K_{\pi_{t}(j)}^{t}+K_{\pi_{s}(j)}^{s}\Big)}\Bigg)^{2}\right\}}.\\
\end{aligned}
\end{equation}
For simplicity, the lower bounds in (\ref{48}), (\ref{49}), and (\ref{71}) are respectively marked by ineq\ref{48}, ineq\ref{49}, and ineq\ref{71}, then let ${LB}=\max\{\rm ineq\ref{48},ineq\ref{49},ineq\ref{71}\}$, namely, $\sum\limits_{t=1}^{N}{I}_{\rho}^{c}(\Phi_{t})\geq LB$.

In \cite{27}, Zhang $et~al.$  provided three lower bounds $LB1$, $LB2$, and $LB3$, and the uncertainty relation $\sum\limits_{t=1}^{N}{I}_{\rho}^{c}(\Phi_{t})\geq\max\{LB1,LB2,LB3\}$ (see reference \cite{27} in detail).

According to the relationship between the norm inequalities given by (\ref{034}), (\ref{035}), and (\ref{62}), it is not hard to show that the result $LB$ derived by us is larger than the lower bound $\max\{LB1,LB2,LB3\}$ in \cite{27}.

The above results (\ref{48}), (\ref{49}), and (\ref{71}) are also satisfied for special cases of metric-adjusted skew information.

Noted that the lower bounds of inequalities (\ref{27}) and (\ref{48}), (\ref{28}) and (\ref{49}), (\ref{68}) and (\ref{71}) are not equal in general. That is to say, the lower bounds obtained by the two distinct expressions of the sum uncertainty relation associated with channels are generally different. Next we will show that the lower bounds in (\ref{48}), (\ref{49}), and (\ref{68}) are greater than the lower bounds in (\ref{27}), (\ref{28}), and (\ref{71}), respectively. Because ${I}_{\rho}^{c}(\cdot)$ is nonnegative, the key is to prove $\Bigg(\sum\limits_{1\leq t<s\leq N} \sqrt{\sum\limits_{j=1}^{n}{I}_{\rho}^{c}\Big(K_{\pi_{t}(j)}^{t}\pm K_{\pi_{s}(j)}^{s}\Big)}\Bigg)^2\geq\sum\limits_{j=1}^{n}\Bigg(\sum\limits_{1\leq t<s\leq N} \sqrt{{I}_{\rho}^{c}\Big(K_{\pi_{t}(j)}^{t}\pm K_{\pi_{s}(j)}^{s}\Big)}\Bigg)^{2}$. If we set $y_{\pi_t(j),\pi_s(j)}^{t,s}={I}_{\rho}^{c}\Big(K_{\pi_{t}(j)}^{t}\pm K_{\pi_{s}(j)}^{s}\Big)$, then $\sum\limits_{j=1}^{n}\Bigg(\sum\limits_{1\leq t<s\leq N} \sqrt{{I}_{\rho}^{c}\Big(K_{\pi_{t}(j)}^{t}\pm K_{\pi_{s}(j)}^{s}\Big)}\Bigg)^{2}=\Big(\sqrt{y_{\pi_1(1),\pi_2(1)}^{1,2}}+\sqrt{y_{\pi_1(1),\pi_3(1)}^{1,3}}+\cdot\cdot\cdot+\sqrt{y_{\pi_{N-1}(1),\pi_{N}(1)}^{N-1,N}}\Big)^2+
\Big(\sqrt{y_{\pi_1(2),\pi_2(2)}^{1,2}}+\sqrt{y_{\pi_1(2),\pi_3(2)}^{1,3}}+\cdot\cdot\cdot+\sqrt{y_{\pi_{N-1}(2),\pi_{N}(2)}^{N-1,N}}\Big)^2+\cdot\cdot\cdot+
\Big(\sqrt{y_{\pi_1(n),\pi_2(n)}^{1,2}}+\sqrt{y_{\pi_1(n),\pi_3(n)}^{1,3}}+\cdot\cdot\cdot+\sqrt{y_{\pi_{N-1}(n),\pi_{N}(n)}^{N-1,N}}\Big)^2$,
$\Bigg(\sum\limits_{1\leq t<s\leq N} \sqrt{\sum\limits_{j=1}^{n}{I}_{\rho}^{c}\Big(K_{\pi_{t}(j)}^{t}\pm K_{\pi_{s}(j)}^{s}\Big)}\Bigg)^2=\Big(\sqrt{y_{\pi_1(1),\pi_2(1)}^{1,2}+y_{\pi_1(2),\pi_2(2)}^{1,2}+\cdot\cdot\cdot+y_{\pi_1(n),\pi_2(n)}^{1,2}}
+\sqrt{y_{\pi_1(1),\pi_3(1)}^{1,3}+y_{\pi_1(2),\pi_3(2)}^{1,3}+\cdot\cdot\cdot+y_{\pi_1(n),\pi_3(n)}^{1,3}}+\cdot\cdot\cdot+
\sqrt{y_{\pi_{N-1}(1),\pi_N(1)}^{N-1,N}+y_{\pi_{N-1}(2),\pi_{N}(2)}^{N-1,N}+\cdot\cdot\cdot+y_{\pi_{N-1}(n),\pi_N(n)}^{N-1,N}}\Big)^2$.
Given some permutation, there are $\frac{nN(N-1)}{2}$ different values here, we suppose the sets $\{y_{j}^{1},y_{j}^{2},\cdot\cdot\cdot,y_{j}^{a}\}$ and $\{y_{\pi_t(j),\pi_s(j)}^{t,s}|1\leq t<s\leq N\}=\{y_{\pi_1(j),\pi_2(j)}^{1,2},y_{\pi_1(j),\pi_3(j)}^{1,3},\cdot\cdot\cdot,y_{\pi_1(j),\pi_N(j)}^{1,N},y_{\pi_2(j),\pi_3(j)}^{2,3},
y_{\pi_2(j),\pi_4(j)}^{2,4},\cdot\cdot\cdot,y_{\pi_2(j),\pi_N(j)}^{2,N},\cdot\cdot\cdot,y_{\pi_N-1(j),\pi_N(j)}^{N-1,N}\}$ correspond in order of elements, $j=1,2,\cdot\cdot\cdot n$, $a=\frac{N(N-1)}{2}$. By subtracting, we derive
$\Big(\sqrt{y_{\pi_1(1),\pi_2(1)}^{1,2}}+\sqrt{y_{\pi_1(1),\pi_3(1)}^{1,3}}+\cdot\cdot\cdot+\sqrt{y_{\pi_{N-1}(1),\pi_{N}(1)}^{N-1,N}}\Big)^2+
\Big(\sqrt{y_{\pi_1(2),\pi_2(2)}^{1,2}}+\sqrt{y_{\pi_1(2),\pi_3(2)}^{1,3}}+\cdot\cdot\cdot+\sqrt{y_{\pi_{N-1}(2),\pi_{N}(2)}^{N-1,N}}\Big)^2+\cdot\cdot\cdot+
\Big(\sqrt{y_{\pi_1(n),\pi_2(n)}^{1,2}}+\sqrt{y_{\pi_1(n),\pi_3(n)}^{1,3}}+\cdot\cdot\cdot+\sqrt{y_{\pi_{N-1}(n),\pi_{N}(n)}^{N-1,N}}\Big)^2
-\Big(\sqrt{y_{\pi_1(1),\pi_2(1)}^{1,2}+y_{\pi_1(2),\pi_2(2)}^{1,2}+\cdot\cdot\cdot+y_{\pi_1(n),\pi_2(n)}^{1,2}}
+\sqrt{y_{\pi_1(1),\pi_3(1)}^{1,3}+y_{\pi_1(2),\pi_3(2)}^{1,3}+\cdot\cdot\cdot+y_{\pi_1(n),\pi_3(n)}^{1,3}}+\cdot\cdot\cdot+
\sqrt{y_{\pi_{N-1}(1),\pi_N(1)}^{N-1,N}+y_{\pi_{N-1}(2),\pi_{N}(2)}^{N-1,N}+\cdot\cdot\cdot+y_{\pi_{N-1}(n),\pi_N(n)}^{N-1,N}}\Big)^2
=\big(\sqrt{y_{1}^{1}}+\sqrt{y_{1}^{2}}+\cdot\cdot\cdot+\sqrt{y_{1}^{a}}\big)^2+
\big(\sqrt{y_{2}^{1}}+\sqrt{y_{2}^{2}}+\cdot\cdot\cdot+\sqrt{y_{2}^{a}}\big)^2+\cdot\cdot\cdot+
\big(\sqrt{y_{n}^{1}}+\sqrt{y_{n}^{2}}+\cdot\cdot\cdot+\sqrt{y_{n}^{a}}\big)^2
-\big(\sqrt{y_{1}^{1}+y_{2}^{1}+\cdot\cdot\cdot+y_{n}^{1}}
+\sqrt{y_{1}^{2}+y_{2}^{2}+\cdot\cdot\cdot+y_{n}^{2}}+\cdot\cdot\cdot+
\sqrt{y_{1}^{a}+y_{2}^{a}+\cdot\cdot\cdot+y_{n}^{a}}\big)^2
=2\sum\limits_{p<q}\big(\sqrt{y_{1}^{p }y_{1}^{q}}+\sqrt{y_{2}^{p}y_{2}^{q}}+\cdot\cdot\cdot+\sqrt{y_{n}^{p}y_{n}^{q}}\big)-2\sum\limits_{p<q}\sqrt{y_{1 }^{p}+y_{2}^{p}+\cdot\cdot\cdot+y_{n}^{p}}\sqrt{y_{1}^{q}+y_{2}^{q}+\cdot\cdot\cdot+y_{n}^{q}}
\leq0$, where the inequality is obtained because $\big(\sqrt{y_{1}^{p }y_{1}^{q}}+\sqrt{y_{2}^{p}y_{2}^{q}}+\cdot\cdot\cdot+\sqrt{y_{n}^{p}y_{n}^{q}}\big)^2\leq({y_{1 }^{p}+y_{2}^{p}+\cdot\cdot\cdot+y_{n}^{p}})({y_{1}^{q}+y_{2}^{q}+\cdot\cdot\cdot+y_{n}^{q}})$ holds based on the triangle inequality. Due to the arbitrariness of permutation, the above conclusion holds for every permutation. The proof completes.

It is obvious that $\max\{\rm ineq\ref{48}, ineq\ref{49}, ineq\ref{68}\}$ is more precise than the lower bounds $\max\{\rm ineq\ref{27},ineq\ref{28},ineq\ref{68}\}$ and $\max\{\rm ineq\ref{48}, ineq\ref{49}, ineq\ref{71}\}$. Therefore, we can get
\begin{equation}\label{36}
\sum\limits_{t=1}^{N}{I}_{\rho}^{c}(\Phi_{t})\geq\max\{\rm ineq\ref{48}, ineq\ref{49}, ineq\ref{68}\}.\\
\end{equation}
For simplicity, the right side of inequality (\ref{36}) is marked by $\overline{LB}$.

To illustrate the tightness of our results, we compare the results obtained by us with existing results. The following we will show two examples based on Wigner-Yanase-Dyson skew information where we take $m=2$, $l=1$ for inequality (\ref{48}), and $m=1$, $l=2$ for inequalities (\ref{68}) and (\ref{49}). One is that each channel has the same number of Kraus operators, and the other is that each channel has a different number of Kraus operators.

\begin{figure}[htbp]
\centering
\subfigure[]{\includegraphics[width=8cm,height=5cm]{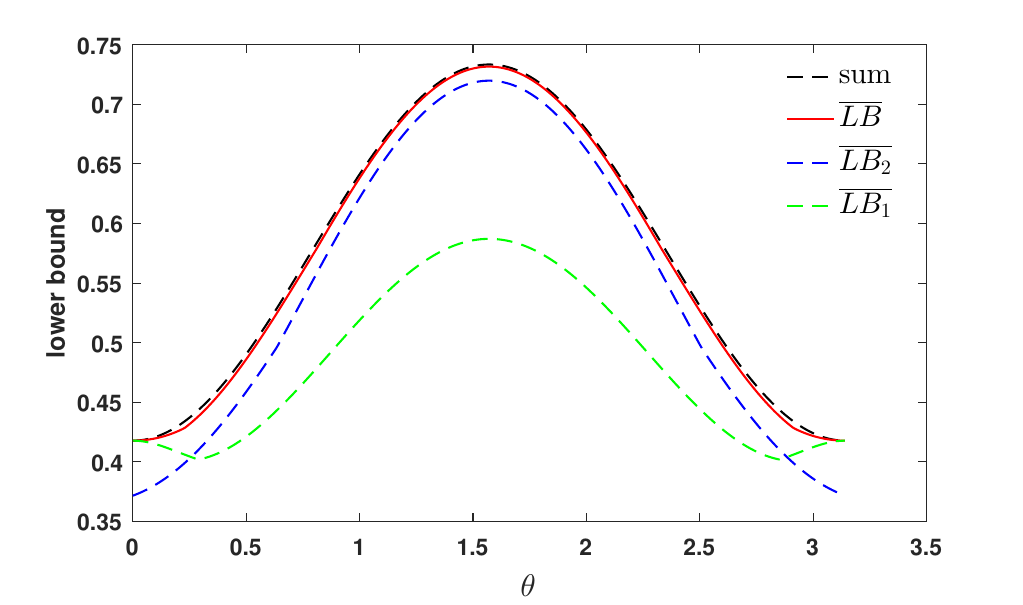}}
\quad
\subfigure[]{\includegraphics[width=8cm,height=5cm]{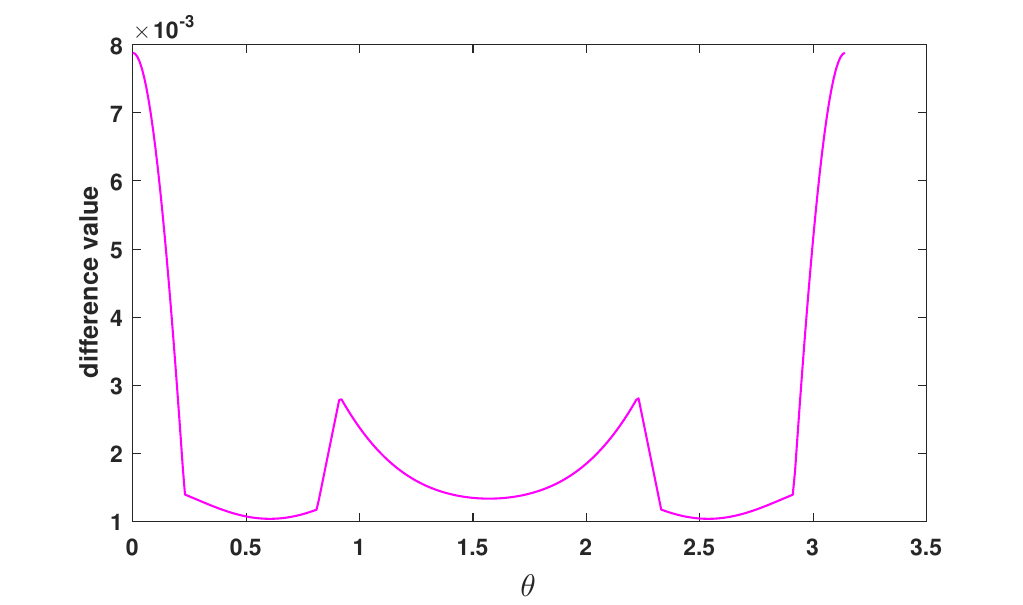}}
\caption{Set $\alpha=\frac{1}{3}$ and $\gamma=0.7$. The black dashed line expresses the value of $I_{\rho}^{1/3}(\Lambda)+I_{\rho}^{1/3}(\varepsilon)+I_{\rho}^{1/3}(\phi)$; the red line, the blue dashed line, and the green dashed line represent the lower bounds $\overline{LB}$, $\overline{LB_{2}}$ and $\overline{LB_{1}}$, respectively. Obviously, the value of $\overline{LB}$ is always larger than $\overline{LB_{2}}$ and $\overline{LB_{1}}$. The (b) shows the difference value between the lower bound $\overline{LB}$ and the lower bound in \cite{27}, namely, $\overline{LB}-\max\{LB1,LB2,LB3\}>0$.} \label{fig 4}
\end{figure}

\textbf{Example 3}. Assume a mixed state $\rho=\frac{I+\vec{r}\cdot\vec{\sigma}}{2}$ with $\vec{r}$=($\frac{\sqrt{3}}{2}$cos$\theta$, $\frac{\sqrt{3}}{2}$sin$\theta$, 0), $0\leq\theta\leq\pi$, and three channels
$\Lambda(\rho)=\sum\limits_{j=1}^{2}E_{j}\rho E_{j}^{\dagger}$~with~$E_{1}=\sqrt{1-\gamma}(|0\rangle\langle0|+|1\rangle\langle1|)$,~ $E_{2}=\sqrt{\gamma}(|0\rangle\langle1|+|1\rangle\langle0|)$,
$\varepsilon(\rho)=\sum\limits_{j=1}^{2}F_{j}\rho F_{j}^{\dagger}$~with~$F_{1}=\sqrt{1-\gamma}(|0\rangle\langle0|+|1\rangle\langle1|)$,~ $F_{2}=\sqrt{\gamma}(|0\rangle\langle0|-|1\rangle\langle1|)$,
$\phi(\rho)=\sum\limits_{j=1}^{2}K_{j}\rho K_{j}^{\dagger}$~with~$K_{1}=|0\rangle\langle0|+\sqrt{1-\gamma}|1\rangle\langle1|$,~ $K_{2}=\sqrt{\gamma}|1\rangle\langle1|$,
are called bit-flipping channel $\Lambda$, phase-flipping channel $\varepsilon$, and amplitude damping channel $\phi$, respectively,  where $0\leq \gamma\leq1$. Then according to (\ref{25}) and (\ref{26}), one has ${I}_{\rho}^{\alpha}(\Lambda)+{I}_{\rho}^{\alpha}(\varepsilon)+{I}_{\rho}^{\alpha}(\phi)\geq{\rm{max}}\left\{A_{1},A_{2},A_{3},A_{4}\right\}$ and ${I}_{\rho}^{\alpha}(\Lambda)+{I}_{\rho}^{\alpha}(\varepsilon)+{I}_{\rho}^{\alpha}(\phi)\geq{\rm{max}}\left\{B_{1},B_{2},B_{3},B_{4}\right\}$, where
$A_{j},~B_{j}$ $(j=1,2,3,4)$ are the lower bounds corresponding to $\left\{\pi_{1}=(1), \pi_{2}=(1), \pi_{3}=(1)\right\}$, $\left\{\pi_{1}=(1), \pi_{2}=(12), \pi_{3}=(12)\right\}$, $\left\{\pi_{1}=(1), \pi_{2}=(1), \pi_{3}=(12)\right\}$ or $\left\{\pi_{1}=(1), \pi_{2}=(12), \pi_{3}=(1)\right\}$.
Analogously, one can get ${I}_{\rho}^{\alpha}(\Lambda)+{I}_{\rho}^{\alpha}(\varepsilon)+{I}_{\rho}^{\alpha}(\phi)\geq\{\max \{C_{1},C_{2},C_{3},C_{4}\},\max\{D_{1},D_{2},D_{3},D_{4}\},\max\{N_{1},N_{2},N_{3},N_{4}\}\}$ by inequalities (\ref{48}), (\ref{49}), and (\ref{68}), respectively. Here the lower bounds $C_{j},~D_{j},~N_{j}$ are similar to $A_{j},~B_{j}$, where $j=1,2,3,4$. Here $\pi_{1}$, $\pi_{2}$, $\pi_{3}$ need to take all of the binary permutations, but the lower bounds in the case $\left\{\pi_{1}=(1), \pi_{2}=(1), \pi_{3}=(1)\right\}$ and the case $\left\{\pi_{1}=(12), \pi_{2}=(12), \pi_{3}=(12)\right\}$ are same, similarly the lower bounds in the cases $\left\{\pi_{1}=(1), \pi_{2}=(12), \pi_{3}=(12)\right\}$ and $\left\{\pi_{1}=(12), \pi_{2}=(1), \pi_{3}=(1)\right\}$, $\left\{\pi_{1}=(1), \pi_{2}=(1), \pi_{3}=(12)\right\}$ and $\left\{\pi_{1}=(12), \pi_{2}=(12), \pi_{3}=(1)\right\}$, $\left\{\pi_{1}=(1), \pi_{2}=(12), \pi_{3}=(1)\right\}$ and $\left\{\pi_{1}=(12), \pi_{2}=(1), \pi_{3}=(12)\right\}$ are same, so we only need to consider four cases. When $\alpha=\frac{1}{3}$ and $\gamma=0.7$, apparently, the lower bound $\overline{LB}$ we had is always greater than the lower bounds $\overline{LB_{2}}$ and $\overline{LB_{1}}$, and our result $\overline{LB}$ is highly close to $I_{\rho}^{1/3}(\Lambda)+I_{\rho}^{1/3}(\varepsilon)+I_{\rho}^{1/3}(\phi)$, which is illustrated in FIG. \ref{fig 4}(a). The FIG. \ref{fig 4}(b) shows that the lower bound $\overline{LB}$ is greater than the lower bound $\max\{LB1,LB2,LB3\}$ in \cite{27}.

\begin{figure}[htbp]
\centering
\subfigure[]{\includegraphics[width=8cm,height=5cm]{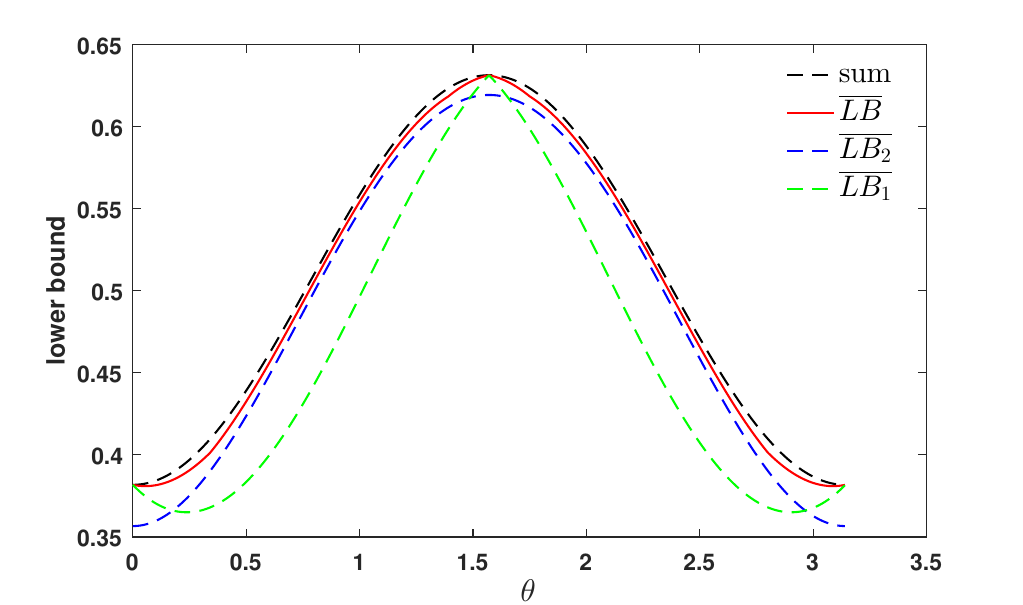}}
\quad
\subfigure[]{\includegraphics[width=8cm,height=5cm]{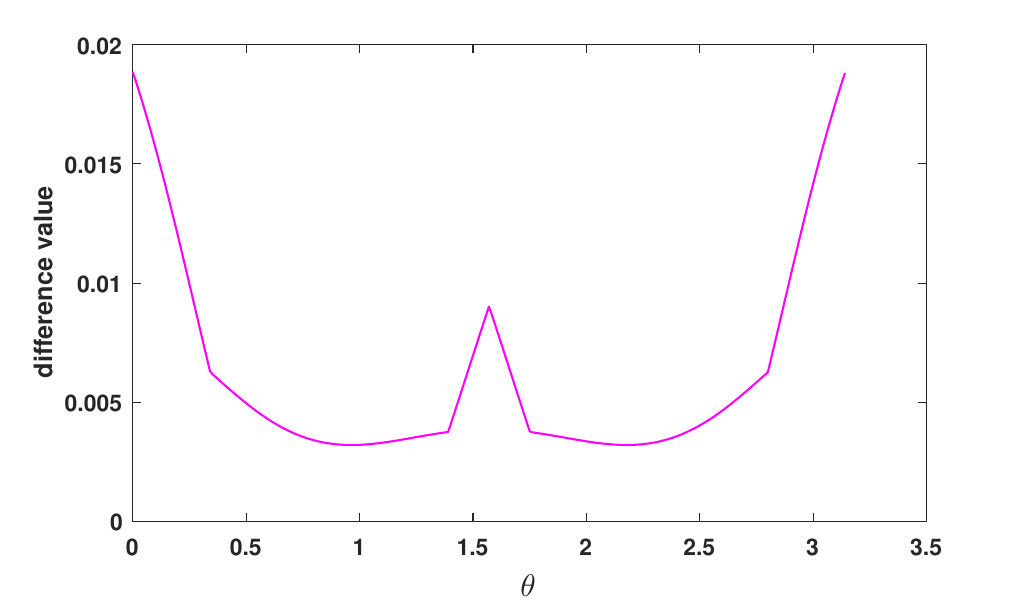}}
\caption{Set $\alpha=\frac{1}{3}$ and $\gamma=0.7$. In (a), the black dashed line is the value of $I_{\rho}^{1/3}(\Lambda)+I_{\rho}^{1/3}(\varepsilon)+I_{\rho}^{1/3}(U)$; the red line, the blue dashed line, and the green dashed line represent the lower bounds $\overline{LB}$, $\overline{LB_{2}}$ and $\overline{LB_{1}}$, respectively. Obviously, the lower bound $\overline{LB}$ is always tighter than $\overline{LB_{1}}$ and $\overline{LB_2}$.  The (b) shows the difference value between the lower bound $\overline{LB}$ and the lower bound in \cite{27}, namely, $\overline{LB}-\max\{LB1,LB2,LB3\}>0$.} \label{fig 5}
\end{figure}

\textbf{Example 4}. Assume that the chosen quantum state is the same as in Example 3, we consider three channels here which are bit-flipping channel $\Lambda$, phase-flipping channel $\varepsilon$, and one unitary channel $U$, respectively, where $\Lambda(\rho)=\sum\limits_{j=1}^{2}E_{j}\rho E_{j}^{\dagger}$ with $E_{1}=\sqrt{1-\gamma}(|0\rangle\langle0|+|1\rangle\langle1|)$, $E_{2}=\sqrt{\gamma}(|0\rangle\langle1|+|1\rangle\langle0|)$, $\varepsilon(\rho)=\sum\limits_{j=1}^{2}F_{j}\rho F_{j}^{\dagger}$ with $F_{1}=\sqrt{1-\gamma}(|0\rangle\langle0|+|1\rangle\langle1|)$, $F_{2}=\sqrt{\gamma}(|0\rangle\langle0|-|1\rangle\langle1|)$, $0\leq \gamma\leq1$, and $U={\rm cos}{\frac{\pi}{8}|0\rangle\langle0|}+{\rm sin}{\frac{\pi}{8}|0\rangle\langle1|}-{\rm sin}{\frac{\pi}{8}|1\rangle\langle0|}+{\rm cos}{\frac{\pi}{8}|1\rangle\langle1|}$. Since each channel has a different number of Kraus operators, we adopt the method of supplementing $\mathbf{0}$ proposed by Ren $et~al$. in \cite{1}. We then use the same procedure as in Example 3. When $\alpha=\frac{1}{3}$ and $\gamma=0.7$, one can see that $\overline{LB}$ is stronger than $\overline{LB_{1}}$ and $\overline{LB_{2}}$, which is illustrated in FIG. \ref{fig 5}(a). Compared the lower bound $\overline{LB}$ with the lower bound $\max\{LB1,LB2,LB3\}$ in \cite{27} as shown in FIG. \ref{fig 5}(b), the result $\overline{LB}$ is larger.

\section{Conclusion}\label{V}
To sum up, we have obtained the new sum uncertainty relations with regard to metric-adjusted skew information of any finite observables and quantum channels by means of the norm inequalities we constructed, and proved our results are stronger than some results in \cite{3,1,27}. The results also definitely hold for its special measures, and we have shown that our results are stronger than some results in \cite{2,4,29} with respect to Wigner-Yanase skew information. For the two different uncertainty relations of channels, when utilizing the norm inequality (\ref{60}), the lower bound derived directly by first form is better, when using the norm inequality (\ref{013}) and (\ref{014}), the results yielded by second form are superior. Using this result we gave an optimal bound. Meanwhile, several specific examples were given to illustrate more clearly that the conclusions we have drawn are superior to the lower bounds in \cite{3,1,27}. We think by using the general form of Lemma 1, one can obtain much better result. It is hoped that our results can provide some reference for further research on sum uncertainty relations.

\section*{ACKNOWLEDGMENTS}
This work was supported by the National Natural Science Foundation of China under Grant No. 12071110, the Hebei Natural Science Foundation of China under Grant No. A2020205014, and funded by Science and Technology Project of Hebei Education Department under Grant Nos. ZD2020167, ZD2021066.


\end{document}